\documentclass[a4paper,11pt]{article}
\usepackage{jheppub} % for details on the use of the package, please see the JINST-author-manual
\allowdisplaybreaks[4]
\title{Fluctuation-dissipation relation in cosmic microwave background}

\author{Atsuhisa Ota}
\affiliation{Department of Physics and Chongqing Key Laboratory for Strongly Coupled Physics, \\
Chongqing University, Chongqing 401331, People's Republic of China}

\emailAdd{aota@cqu.edu.cn}

\abstract{
We study the fluctuation-dissipation relation for sound waves in the cosmic microwave background (CMB), employing effective field theory (EFT) for fluctuating hydrodynamics. 
Treating sound waves as the linear response to thermal radiation, we establish the fluctuation-dissipation relation within a cosmological framework. 
While dissipation is elucidated in established linear cosmological perturbation theory, the standard Boltzmann theory overlooks the associated noise, possibly contributing to inconsistencies in Lambda Cold Dark Matter ($\Lambda$CDM) cosmology. This paper employs EFT for fluctuating hydrodynamics in cosmological perturbation theory, deriving sound wave noise. 
Notably, the long-time limit of the noise spectrum is independent of viscosity details, resembling a Brownian motion bounded in a harmonic potential. 
The net energy transfer between the sound wave system and the radiation environment reaches a balance within Hubble time, suggesting the thermal equilibrium of the sound waves themselves. 
The induced density power spectrum is characterized as white noise dependent on the inverse of the entropy density, which is negligibly small on the CMB scale.  The energy density of the entire sound wave system scales as $a^{-4}$, akin to radiation. While the numerical factor is not determined in the present calculation, the back reaction of the sound wave system to the background radiation may not be negligible, serving as a potential source for various fitting issues in $\Lambda$CDM cosmology.

}

\begin{document}
\maketitle
\flushbottom

\section{Introduction}

The fluctuation-dissipation relation is a universal property of the linear response to thermal systems~\cite{Kubo:1966fyg}. 
For example, it applies to a particle within a thermal molecular environment. 
A fundamental relationship exists between the friction and random driving force on the Brownian particle.

In cosmology, our focus lies on the propagation of sound waves in the cosmic microwave background (CMB).
The sound wave propagating on the CMB last scattering surface is seen as an anisotropy in the present sky. 
We estimate cosmological parameters by fitting the CMB anisotropies, putting forward a cosmological model~\cite{WMAP:2003elm,Planck:2015fie}. 
The standard general relativistic Boltzmann theory explains the dissipation of the sound waves and thus the observed damping feature in the CMB~(see, e.g., Ref.~\cite{Dodelson:2003ft} and references therein.).
However, unlike Brownian motion, random fluctuation in the sound waves is absent in the kinetic theory.
This is because the equations of motion in the Boltzmann theory are classical and do not incorporate both classical statistical and quantum noise. 
This randomness is distinct from the random initial conditions of super-horizon cosmological perturbations set by inflation~\cite{Dodelson:2003ft}. 
One may regard these sound waves as the hydrodynamical modes within the thermal radiation environment. 
If the fluctuation-dissipation relation applies to the CMB, similar to other thermal systems, the random noise is characterized by temperature and exists irrespective of initial conditions.
That is, the CMB may be inherently noisy.

Recently, challenges to the $\Lambda$ cold dark matter~(CDM) cosmology have surfaced as observational techniques advance and data precision improves~\cite{DiValentino:2021izs,Perivolaropoulos:2021jda}. 
Discrepancies among cosmological observables or inconsistencies within the same measurements suggest missing elements. 
Potential sources include improper treatment of experimental errors or the necessity of incorporating new physics into the $\Lambda$CDM model.

This paper delves into the latter case, specifically examining the random noise in sound waves when applying the fluctuation-dissipation relation to the CMB. 
Intuitively, on the analogy of the linear response to other thermal systems, random noise should arise during diffusion, suggesting scale-dependent noise, potentially impacting the interpretation of the temperature angular power spectrum.
Despite over three decades of cosmological perturbation theory, this extension is novel to the best of our knowledge. 
In contrast to other approaches, ours is new but does not rely on unknown physics, as the fluctuation-dissipation relation is universal.

Our analysis is limited to the tree-level and semiclassical.
Hence, one approach is to employ the traditional formulation of hydrodynamical fluctuations with local Gaussian distributions~(see, e.g.,~Ref.~\cite{Kovtun:2012rj} for a comprehensive review.).
However, our goal is to establish the fluctuation-dissipation relation in a general setup in cosmological perturbation theory and apply it to the CMB.
Considering applications to curved spacetime and various cosmological setups, an EFT based approach with symmetries and action principle will be more advantageous than the phenomenological methods.
Action formulation for hydrodynamics has a long history, indeed, and a breakthrough emerges with recent progress in the effective field theory of dissipative fluids~(see Refs~\cite{Crossley:2015evo,Glorioso:2017fpd} and references therein.). 
By identifying the ``hydrodynamical mode'' and its effective action, they constructed the effective theory of fluctuating hydrodynamics from symmetry principles.
An advantage of the EFT is that fundamental symmetries constrain the EFT coefficients automatically.
The fluctuation-dissipation relation, or the Kubo-Martin-Schwinger~(KMS) condition~\cite{Kubo:1957mj,Martin:1959jp,Kadanoff:1963axw} for correlation functions, can be reduced to the invariance under a $Z_{2}$ symmetry—referred to as KMS symmetry—of the effective action. 
In the present case, we can reconstruct the classical hydrodynamical action, referring to the output of the Boltzmann theory.
Then, we may extend the hydrodynamical action from the classical to the semiclassical regime by imposing the invariance under the KMS symmetry.
The random noise is described in the semiclassical limit of the effective action, which enables us to identify the magnitude of noise when a fluctuation-dissipation relation is established. 

This paper is organized as follows.
Section~\ref{review} is a review of the damping feature in sound waves in the CMB.
We compute the dispersion relation for the sound waves in the CMB by using the Boltzmann theory.
In Section~\ref{eftf}, we formulate the EFT of fluctuating hydrodynamics.
We present all necessary theoretical details concisely.
Then, we apply the EFT to the cosmological setup in Section~\ref{sec:cosmofluid}.
We consider the EFT of fluctuating hydrodynamics in a perturbed Friedmann Lemaitre Robertson Walker~(FLRW) spacetime and derive the noise spectrum in cosmological perturbation theory without specifying fluids.
In Section~\ref{pbp}, we consider photon-baryon plasma as a specific example of cosmological fluids.
In the deep radiation-dominant universe, many EFT coefficients are simplified, and we estimate the size of the noise.
Section~\ref{eib} considers the energy stored in the sound wave noise and discusses the backreaction to the background spacetime.
We conclude this paper in Section \ref{conc}.
Appendix~\ref{apti} is a summary of a mathematical theorem used to show the unitarity constraints in the Schwinger-Keldysh formalism, which was implicit in the literature.
We derive the dynamical KMS transformation for the hydrodynamical variables in Appendix~\ref{apdy}.

\section{Sound waves in the CMB}
\label{review}

%Before last scattering of the CMB, the universe was filled with the photon-baryon plasma whose fluctuations are the seeds of present cosmological large scale structure.
%The dynamics of the tiny perturbations are described by the linearized Einstein equation.
%We put the energy-momentum tensor of cosmological fluids, such as radiation and dark matter, in the Einstein equation.
%We then expand the equation of motion to first order in cosmological perturbations and find the evolution of the inhomogeneity.
%The evolution of cosmological structure is considered in cosmological perturbation theory.

Sound waves in the CMB are described in the linearized Boltzmann-Einstein theory in a flat FLRW background.
In Fourier space, the linearized continuity equation and Euler equation for photons are given as~\cite{Ma:1995ey} 
\begin{align}
	\dot \delta_\gamma{}& =-\frac{4}{3}\theta_\gamma -\dot \psi  ,\label{cont:eq}
	\\
	\dot \theta_\gamma &= k^2 \left( \frac{1}{4}\delta_\gamma - \Sigma_\gamma\right) + k^2 \phi+ \dot \tau \theta_{\gamma b}, \label{eul:eq}
\end{align}
where $\delta_{\gamma}$, and $\theta_{\gamma}$ are the density fluctuation and velocity divergence of radiation.
$\phi$ and $\psi$ are the gravitational potential and curvature fluctuation in the conformal Newtonian gauge~(see Section~\ref{sec:cosmofluid} for more details.).
The over-dot is a derivative with respect to the conformal time, and $k$ is the magnitude of the Fourier wavenumber.
The differential optical depth $\dot \tau$ is written as
\begin{align}
		\dot \tau = -n_{\rm e}\sigma_{\rm T}a,\label{def:dottau}
\end{align}
where $n_{\rm e}$, $\sigma_{\rm T}$ and $a$ are the number density of electrons, the Thomson scattering cross-section, and the scale factor.
$\dot \tau<0$ implies that the number of scattering events decreases as the the universe expands.
Eqs.~\eqref{cont:eq} and \eqref{eul:eq} are not closed for $\delta_{\gamma}$, and $\theta_{\gamma}$ because of the photon-baryon slip $\theta_{\gamma b}$ and shear $\Sigma_\gamma$.
These are determined in the higher-order hydrodynamical equations in the gradient expansion, and the whole set of equations is called the Boltzmann hierarchy.
In practice, we solve the background Friedmann equation and the recombination history to find $\tau$.
Combining these background solutions with the Boltzmann hierarchy and the linearized Einstein equation, one can solve Eqs.~\eqref{cont:eq} and \eqref{eul:eq}.
Here, for simplicity, we truncate the Boltzmann hierarchy at the leading order and ignore the metric perturbations in the sub-horizon limit.
In this case, using $\delta_\gamma$ and $\theta_\gamma$, $\theta_{\gamma b}$ and $\Sigma_\gamma$ are roughly written as~\cite{Dodelson:2003ft}
\begin{align}
	\theta_{\gamma b} &\approx -\theta_\gamma \left( 
	\left(\frac{ic_sk R}{\dot \tau}\right)	
	+
	\left(\frac{ic_sk R}{\dot \tau}\right)^2
	\right),\label{aprox:bg}
\\
	\Sigma_\gamma &\approx- \frac{8}{27 \dot \tau}\theta_\gamma,\label{aprox:sh}
\end{align}
where $c_s$ is the sound speed defined as
\begin{align}
		c_s \equiv \frac{1}{\sqrt{3(1+R)}},~R\equiv \frac{3\bar \rho_b}{4\bar \rho_\gamma},\label{defR}
\end{align}
with the background energy density of baryons and photons, $\bar \rho_b$ and $\bar \rho_\gamma$.
Substituting Eqs.~\eqref{aprox:bg} and \eqref{aprox:sh} into Eqs.~\eqref{cont:eq} and \eqref{eul:eq}, we find the dispersion relation for the sound wave:
\begin{align}
	{\rm Re} \omega &= c_s k ,
	\\
	{\rm Im} \omega &\approx -\frac{k^2}{2(1+R)\dot \tau}\left(c_{s}^2 R^2 + \frac{8}{27}\right).\label{dispersion}
\end{align}
The sound wave is dissipative as ${\rm Im \omega}>0 $ as the WKB solution is written as $\sim e^{i\omega t}$.
We see the damping feature in the CMB anisotropies.
Thus, the dissipative feature is explained in the standard linear cosmological perturbation theory based on the kinetic theory.

The sound wave is the linear response to the thermal CMB, which should follow the fluctuation-dissipation relation.
On the analogy of Brownian motion in a thermal environment, we anticipate the scale-dependent noise $\xi$ in the RHS of Eqs.~\eqref{cont:eq} and \eqref{eul:eq} whose statistical property should be written as
\begin{align}
	\langle \xi_{\mathbf k}(x^0)\xi_{\bar{\mathbf k}}(\bar x^0)\rangle = (2\pi)^3 \delta^{(3)}(\mathbf k+\bar{\mathbf k})\delta(x^0-\bar x^0)A_k(x^0).\label{2ptanticipate} 
\end{align}
$A_k$ will be found from the fluctuation-dissipation relation and the dissipative coefficient in Eq.~\eqref{dispersion}.
In the presence of stochastic noise, the CMB can be inherently noisy even if the initial conditions are zero.

In thermal quantum field theory, the fluctuation-dissipation relation can be found as a relationship among the three propagators, i.e., statistical, retarded, and advanced propagators, which is known as the KMS condition~\cite{Kubo:1957mj,Martin:1959jp,Kadanoff:1963axw}.
In that case, dynamical variables are the microscopic fields, and then the calculation of the correlation functions is straightforward once we identify the classical action and an initial thermal state~(see, e.g., Ref.~\cite{Bellac:2011kqa}).
Sound waves are macroscopic dynamical variables that emerge from an underlying microscopic theory.
Hence, the action formalism is not straightforward.
We must carefully review the symmetries inherent in the hydrodynamical action of the sound waves to find the fluctuation-dissipation relation.

\section{Effective theory for hydrodynamics}
\label{eftf}

Long story short, the stochastic noise of sound waves can be described in the effective field theory of fluctuating hydrodynamics~\cite{Crossley:2015evo,Glorioso:2017fpd} discussed in this section.
Readers who are only interested in the cosmological application may skip this section.

In the effective field theory of fluctuating hydrodynamics, we are interested in the correlation functions of the hydrodynamical conserved currents in the presence of stochastic noise.
For example, we want to compute correlation functions of the energy-momentum tensor:
\begin{align}
	\langle T^{\mu\nu} \cdots \rangle.\label{corrTmunu}
\end{align}
%In this paper, we are interested in a neutral fluid as the CMB is not charged.
Given a theory for a microscopic field $\phi$, the classical energy-momentum tensor is defined as the functional derivative of the classical matter action $S[g,\phi]$ with respect to the metric $g$:
\begin{align}
	T^{\mu\nu} = \frac{2}{\sqrt{-\det g}}\frac{\delta S[g,\phi]}{\delta g_{\mu\nu}},\label{tmunudef}
\end{align}
where $\det g$ is the determinant of the metric.
For simplicity, we only consider neutral currents, as the CMB is not charged.
%Thus, the metric enters the classical matter action as an external field, and its functional derivative yields the energy-momentum tensor.
%In the present case, w
%We are interested in the average of this energy-momentum tensor in the presence of the stochastic noise.
Then, the expectation value of $T^{\mu\nu}$ in the presence of the stochastic noise will be described in the path integral in the Schwinger-Keldysh formalism:
\begin{align}
	e^{W[g_{(1)},g_{(2)}]} = \int_{\varrho} D\phi_{(1)} D \phi_{(2)} e^{\frac{i}{\hbar}S[\phi_{(1)}, g_{(1)}]-\frac{i}{\hbar}S[\phi_{(2)}, g_{(2)}]},\label{fullgene}
\end{align}
where $\hbar$ is the reduced Planck constant, and the dynamical variables and external source fields are doubled on the closed time path~(CPT).
$\varrho$ is a density operator, i.e., the initial condition of the path integral.
Given a path integral, Eq.~\eqref{tmunudef} yields
\begin{align}
	\langle T_{(s)}^{\mu\nu} \rangle =  \left. \frac{(-1)^{s+1}2}{\sqrt{-\det g_{(s)}}}\frac{\hbar \delta W[g_{(1)},g_{(2)}]}{i\delta g_{(s)\mu\nu}}\right|_{g_{(1)}=g_{(2)}=0}\label{Tmunus}.
\end{align}
Higher-order correlation functions are defined in the same way.
Eq.~\eqref{fullgene} contains full information about the system, and then Eq.~\eqref{corrTmunu} will be computed, in principle. 
However, the actual path integral is always nontrivial.

In the effective field theory of fluctuating hydrodynamics, we consider a different approach.
The full path integral measure is separated into the measures with respect to the fast variable $\psi_{(s)}$ and the slow variable $\pi_{(s)}$:
\begin{align}
	D\phi_{(1)} D \phi_{(2)}  &\propto D\pi_{(1)}D\pi_{(2)} D\psi_{(1)}D\psi_{(2)},\label{changeval}
\end{align}
where we ignored the Jacobian for simplicity.
Then, the partial path integral with respect to the fast variable is identified with the hydrodynamical effective action:
\begin{align}
	e^{W[g_{(1)},g_{(2)}]} 
	=& \int D\pi_{(1)}D\pi_{(2)} e^{\frac{i}{\hbar}I_{\rm eff.}[\pi_{(1)},g_{(1)}; \pi_{(2)},g_{(2)}]} ,\label{partialpathint}
\end{align}
where we defined
\begin{align}
	e^{\frac{i}{\hbar}I_{\rm eff.}[\pi_{(1)},g_{(1)}; \pi_{(2)},g_{(2)}]} &= \int_{\varrho} D\psi_{(1)}D\psi_{(2)} e^{\frac{i}{\hbar}\bar S[\psi_{(1)},\pi_{(1)},g_{(1)}]-\frac{i}{\hbar}\bar S[\psi_{(2)},\pi_{(2)},g_{(2)}]}.\label{fastint}
\end{align}
In the effective field theory of fluctuating hydrodynamics, we directly construct $I_{\rm eff.}$ by imposing several constraints.
The key questions for the effective theory are 
\begin{itemize}
	\item What are the hydrodynamical modes $\pi_{(s)}^\mu$?
	\item What are the symmetries for $I_{\rm eff.}$?
	\item What are the constraints to define a ``fluid''? 
\end{itemize}

The goal of this section is to write the sound waves' noise for a given classical energy-momentum tensor.
We review the symmetries of hydrodynamics and how to identify the hydrodynamical mode and construct the effective action.
This section is based on Refs.~\cite{Crossley:2015evo}, and we summarize the key components of the framework, and the presentations are tailored for the current application.
We particularly focus on the formulation in the physical spacetime for application to cosmology.
While they also presented the physical spacetime formulation, the fluid spacetime formulation was interchangeably referred to.
Here, we are picky about the physical spacetime and regard the fluid spacetime symmetries as a gauge symmetry in the internal space.

\subsection{Constraints on Schwinger-Keldysh path integrals}
Before identifying the hydrodynamical action, let us summarize general constraints on the Schwinger-Keldysh path integral, which the hydrodynamical effective theory should satisfy.
\begin{enumerate}
\item [1.] Conjugate condition of the density operator: 
\begin{align}
	W^*[g_{(1)},g_{(2)}] &= W[g_{(2)},g_{(1)}].\label{conjugate}
\end{align}
This condition is straightforwardly derived from the operator expression of the path integral:
\begin{align}
	e^{W[g_{(1)},g_{(2)}]} = {\rm Tr}\left[\hat \varrho \hat U^\dagger[g_{(2)}] \hat U[g_{(1)}]\right],\label{oppath}
\end{align}
where $\hat U[g_{(s)}]$ is the time evolution operator from the initial time to the final time, in the presence of the external field $g_{(s)}$.
$\hat \varrho = \hat \varrho^\dagger$ and the trace cyclicity for Eq~\eqref{oppath} lead to condition~\eqref{conjugate}.
	\item [2.] Unitarity condition: 
\begin{align}
	W[g,g] &=0.\label{Wgg=0}
\end{align}
Inserting the same external source to the unitary operators in Eq.~\eqref{oppath}, ${\rm Tr}[\hat \varrho]=1$ yields Eq.~\eqref{Wgg=0}.
Eventually, there is another constraint from unitarity.
Using the trace inequality~\eqref{myeq}, we immediately find 
\begin{align}
	|e^{W[g_{(1)},g_{(2)}]}| \leq 1,
\end{align} 
which implies
\begin{align}
	{\rm Re}~W[g_{(1)},g_{(2)}]\leq 0.\label{uni2}
\end{align}
	\item [3.] The local KMS condition. Consider an initial thermal state written as
\begin{align}
	\hat \varrho_{\bar \beta} \equiv \frac{e^{-\bar \beta \hat H}}{Z},~Z \equiv {\rm Tr}[e^{-\bar \beta \hat H}], \label{def:rhobeta0}
\end{align}
where $\hat H$ and $\bar \beta$ are the initial Hamiltonian and temperature.
$\bar \beta$ is understood as an imaginary time interval.
Hence, we introduce the imaginary time translation $g\to g^{(\bar \beta)}$ by
\begin{align}
		\hat U[g^{(\bar \beta)}] &=e^{-\bar \beta \hat H} \hat U[g] e^{\bar \beta \hat H}.\label{imtrns}
\end{align}
Then, the path integral satisfies
\begin{align}
	W[g_{(1)}, g_{(2)}] &=\frac{1}{Z_0} {\rm Tr}\left[e^{-\bar \beta \hat H} \hat U^\dagger[g_{(2)}]e^{\bar \beta \hat H} e^{-\bar \beta \hat H}\hat U[g_{(1)}]\right]
	\notag 
	\\
	&={\rm Tr}\left[\hat \varrho_{\bar \beta}\hat U[g_{(1)}] \hat U^\dagger[g_{(2)}^{(\bar \beta)}] \right]
	\notag 
	\\
	&\equiv W_{\rm T}[g_{(1)},g_{(2)}^{(\bar \beta)}].\label{deform}
\end{align}
Now we assume the time-reversal symmetry in the path integral:
\begin{align}
	W[g_{(1)},g_{(2)}] = W_{\rm T}[\tilde g_{(1)},\tilde g_{(2)}],\label{timerev}
\end{align}
where $g_{(s)}\to \tilde g_{(s)}$ is the time reversal transformation for the metrics.
Then, Eqs.~\eqref{deform} and \eqref{timerev} yield the KMS condition
\begin{align}
	W[g_{(1)}, g_{(2)}] = W[\tilde g_{(1)}, \tilde g_{(2)}^{(\bar \beta)}],
\end{align}
which implies a $Z_2$ symmetry under the local KMS transformation:
\begin{align}
	g_{(1)} &\to  \tilde  g_{(1)},\label{kmsg1} 
	\\ 
	g_{(2)} &\to   \tilde g_{(2)}^{(\bar \beta)}.\label{kmsg2}
\end{align}
\end{enumerate}

As we mentioned above, these are general requirements for the Schwinger-Keldysh path integral.
The effective field theory of hydrodynamics is not just an effective field theory for the slow variables, as we impose an extra symmetry structure to ``define a fluid''.

\subsection{Keldysh rotation}

Let us introduce a convenient notation in the hydrodynamical effective field theory.
Using the doubled copies on the CTP, we define the Keldysh rotation of the metrics 
\begin{align}
	g_{\mu\nu}(x) &\equiv \frac{g_{(1)\mu\nu}(x)+ g_{(2)\mu\nu}(x)}{2},\label{def:gR}
	\\
	g_{A\mu\nu}(x) &\equiv g_{(1)\mu\nu}(x) - g_{(2)\mu\nu}(x) .\label{def:gA}
\end{align}
The Keldysh rotated energy-momentum tensors are similarly defined as 
\begin{align}
	T_{\mu\nu}(x) &\equiv \frac{T_{(1)\mu\nu}(x)+ T_{(2)\mu\nu}(x)}{2},\label{def:TR}
	\\
	T_{A\mu\nu}(x) &\equiv T_{(1)\mu\nu}(x) - T_{(2)\mu\nu}(x) .\label{def:TA}
\end{align}
Hereafter, we call the $A$-fields ``advanced'' fields. 
Note that the doubled copies have the same configuration for the classical path, which implies the advanced fields are at most $\mathcal O(\hbar)$. 
$T_{(s)}^{\mu\nu}$ are the energy-momentum tensors defined on the outgoing and incoming paths in the CTP, which is not observable.
The observables correspond to the averaged operators.
The CTP doubles the physical spacetime and, thus, the symmetry structure inherent in the system.
In our case, the physical spacetime diffeomorphism is doubled.
Meanwhile, the nondiagonal spacetime diffeomorphism is fixed.
As a result, we only have a physical spacetime manifold, and then the Keldysh rotation is well-defined to all orders in $\hbar$.

\subsection{Physical spacetime symmetry}
From Eq.~\eqref{Tmunus}, the variation of the effective action with respect to the metric is expanded into
\begin{align}
	\delta I_{\rm eff.} &= \sum_{s=1,2}\frac{(-1)^{s+1}}{2} \int d^4 x \sqrt{-g_{(s)}}T_{(s)}^{\mu\nu}\delta g_{(s)\mu\nu}
	\notag
	\\
	&=\frac{1}{2} \int d^4 x \sqrt{-g} ( T^{\mu\nu}_{\rm cl}\delta g_{A\mu\nu}  + T_A^{\mu\nu}\delta g_{\mu\nu}  ) + \mathcal O(\hbar^2),\label{aba}
\end{align}
where we used the Keldysh basis in the last line.
The subscript $\rm cl$ implies the classical component, $T^{\mu\nu} = T^{\mu\nu}_{\rm cl} + \mathcal O(\hbar^2)$.
Then, Eqs.~\eqref{Tmunus} and \eqref{aba} yield
\begin{align}
	\langle T^{\mu\nu}_{\rm cl} \rangle = \lim_{\hbar\to 0} \left. \frac{2}{\sqrt{-\det g}}\frac{\hbar \delta W[g_{(1)},g_{(2)}]}{i \delta g_{A\mu\nu}}\right|_{g_\phi=0}\label{TRmunus},
\end{align}
which implies that, in the lowest order in $\hbar$, the effective action is found as
\begin{align}
	\tilde  I_{1} = \frac{1}{2}\int d^4x  \sqrt{-g} T^{\mu\nu}_{\rm cl}g_{A\mu\nu},\label{def:I1}
\end{align}
where the subscript $1$ implies the order in $\hbar$.
Hereafter, we drop ``cl''.
The conjugate condition and the unitary condition are satisfied for $\tilde I_1$.
The path integral with $\tilde  I_{1}$ is classical since $\hbar$ cancels in the wave function.
Instead, an effective Plank constant with the dimension of $\hbar$ emerges, depending on the details of the system~\cite{Glorioso:2017fpd}.

Next, let us examine the spacetime symmetry structure of the effective action $\tilde I_1$ and identify the hydrodynamical variables $\pi^\mu$ and $\pi_{A\mu}$.
The original Schwinger-Keldysh path integral~\eqref{fullgene} enjoys the invariance under the two independent diffeomorphisms:
\begin{align}
	x^\mu \to \bar x^\mu_{(s)} = x^\mu + \xi^\mu_{(s)}(x).
\end{align}
In the Keldysh basis, the tangents are written as
\begin{align}
	\xi^\mu(x) &\equiv \frac{\xi^\mu_{(1)}(x) + \xi^\mu_{(2)}(x)}{2},\label{def:xiR}
	\\
	\xi^\mu_A(x) &\equiv \xi^\mu_{(1)}(x) - \xi^\mu_{(2)}(x).\label{def:xiA}
\end{align}
As the effective action is defined for a single coordinate system, $x$, it is convenient to impose $\xi^\mu_A(x)=0$, and we will recover the broken diffeomorphism by the Stueckelberg trick.
Under $\xi^\mu$, we find the standard physical spacetime diffeomorphism:
\begin{align}
	g_{\mu\nu}(x) &\to \bar g_{\mu\nu}(\bar x) = \frac{\partial \bar x^\mu}{\partial x^\alpha}\frac{\partial \bar x^\nu}{\partial x^\beta}g_{\alpha \beta}(x),
	\\
	g_{A \mu\nu}(x) &\to \bar g_{A \mu\nu}(\bar x) = \frac{\partial \bar x^\mu}{\partial x^\alpha}\frac{\partial \bar x^\nu}{\partial x^\beta}g_{A \alpha \beta}(x).
\end{align}
The effective action \eqref{def:I1} is invariant under this diagonal diffeomorphism.
The residual spacetime diffeomorphisms $\xi_{(1)}^{\mu} = \xi_A^{\mu}/2$ and $\xi_{(2)}^{\mu} = -\xi_A^{\mu}/2$, yield the following gauge transformation
\begin{align}
	\delta_{\xi_A} g_{(1)\mu\nu}(x) & = \frac{1}{2}\pounds_{\xi_A}  g_{(1)\mu\nu}(x) + \mathcal O(\hbar^2),
	\\
	\delta_{\xi_A} g_{(2)\mu\nu}(x) & = -\frac{1}{2}\pounds_{\xi_A}  g_{(2)\mu\nu}(x)+ \mathcal O(\hbar^2) ,
\end{align}
which can be recast into 
\begin{align}
	\delta_{\xi_A} g_{\mu\nu} &= \mathcal O(\hbar^2),\label{xiAgmunu}
	\\
	\delta_{\xi_A} g_{A \mu\nu} &=\pounds_{\xi_A} g_{\mu\nu}+  \mathcal O(\hbar^3)\label{xiAgmunuA}.
\end{align}
The energy-momentum tensor also transforms in the same way.
With this gauge transformation, Eq.~\eqref{def:I1} transforms as
\begin{align}
	\delta_{\xi_A}\tilde I_1 = \frac{1}{2}\int d^4x \sqrt{-\det g} T^{\mu\nu}\left( \pounds_{\xi_A} g_{\mu\nu} +\mathcal O(\hbar^3)\right).\label{third}
\end{align}
Using the Stueckelberg trick, we introduce the Nambu-Goldstone~(NG) mode as
\begin{align}
	\xi_A^\mu \to -\pi_A^\mu,
\end{align}
and then the residual gauge symmetry of the effective action is recovered as
\begin{align}
	 I_1 = \frac{1}{2}\int d^4x \sqrt{-\det g} T^{\mu\nu}G_{A\mu\nu},
\end{align}
where we defined
\begin{align}
	 G_{A\mu\nu}\equiv   g_{A\mu\nu}  - \pounds_{\pi_A}g_{\mu\nu} =  g_{A\mu\nu} - \nabla_\mu \pi_{A\nu} -\nabla_\nu \pi_{A\mu}.
\end{align}
The gauge transformation of the NG mode is nonlinear:
\begin{align}
	\pi_A^\mu(x) \to \bar \pi_A^\mu(x) = \pi_A^\mu(x) + \xi^\mu_A (x).
\end{align}
$I_1$ is now symmetric under the doubled spacetime diffeomorphisms.
The NG mode $\pi_{A\mu}$ can be regarded as the advanced partner of the hydrodynamical variable that we are looking for.
The equation of motion with respect to $\pi_{A\mu}$ is given as 
\begin{align}
	\frac{1}{\sqrt{-\det g}} \frac{\delta I_1}{\delta \pi_{A\nu}} = \nabla_\mu T^{\mu\nu} = 0,
\end{align}
where we used the Gauss's law:
\begin{align}
	\int d^4x \sqrt{-\det g} T^{\mu\nu} \nabla_\mu \pi_{A\nu} = -\int d^4x \sqrt{-\det g} (\nabla_\mu T^{\mu\nu})  \pi_{A\nu}.
\end{align}
Thus, we introduced the NG mode $\pi_A^\mu$ by restoring the residual gauge symmetry associated with $\xi_A^\mu$.
As a result, the conservation law of the energy-momentum tensor is obtained as the classical equation of motion.
This naturally motivates us to introduce another NG mode by $\xi^\mu\to - \pi^\mu$.
Thus, we introduced the NG modes $\pi^\mu$ and $\pi^\mu_A$ as nonlinear realizations of doubled spacetime diffeomorphisms in the Schwinger-Keldysh formalism.
By construction, $\pi^\mu$ describes the dynamics of the fluid elements, which is identified with the hydrodynamical modes in the effective theory. 
The last question is how to write $T^{\mu\nu}$ with respect to $\pi^\mu$ and how to extend the action to the stochastic regime.

\subsection{Fluid spacetime symmetry}

With the NG mode $\pi^\alpha$, one can introduce four scalar fields
\begin{align}
	\sigma^\alpha(x) = x^\alpha - \pi^\alpha (x). \label{phystofluid}
\end{align}
The four-dimensional internal spacetime characterized by $\sigma^\alpha$ is known as the fluid spacetime whose symmetry structure defines a fluid.
One may define a fluid by imposing the following diffeomorphism invariance in this internal spacetime~\cite{Crossley:2015evo}:
\begin{enumerate}
	\item 
	Time-independent spatial diffeomorphism:
\begin{align}
	\sigma^a \to \bar \sigma^b(\sigma^a).\label{labeling:sp}
\end{align}
	This condition implies that $\sigma^a~(a=1,2,3)$ is a label fixed to a fluid element, and one can define them once at some time.
	\item
	Fluid spacetime-dependent time-diffeomorphism:
	\begin{align}
	\sigma^0 \to \bar \sigma^0(\sigma^0, \sigma^a).\label{labeling:tm}
\end{align}
	$\sigma^0$ is a parameter of individual orbits of the fluid elements, which can be rescaled freely.	
\end{enumerate}
Thus, Eqs.~\eqref{labeling:sp} and \eqref{labeling:tm} are the gauge symmetries to define a fluid by uniquely identifying each fluid element and its orbit.
$T^{\mu\nu}$ will be constructed from the covariant derivative of this scalar: 
\begin{align}
	K_{\mu}{}^\alpha \equiv \nabla_\mu \sigma^\alpha.
\end{align}
Now, the internal spacetime index $\alpha$ appears.
One may consider \eqref{phystofluid} as a coordinate transformation from $x$ to $\sigma$, while $\sigma$ is not a regular coordinate system as it contains the dynamical variables.
With this effective coordinate transformation, we can embed the physical spacetime metric to fluid spacetime as
\begin{align}
	h_{\alpha \beta}(\sigma) &= \frac{\partial x^\mu}{\partial \sigma^\alpha}\frac{\partial x^\nu}{\partial \sigma^\beta}g_{\mu\nu}(x).
\end{align}
$\sigma$ is fixed to a coordinate system at rest with respect to the fluid.
Hence, the fluid velocity is written as
\begin{align}
	u^\alpha(\sigma) = \frac{1}{b} \delta^\alpha{}_0,~b\equiv \sqrt{-h_{00}}.\label{defvalpha}
\end{align}
One can explicitly show that $u^\alpha(\sigma)$ stays in another rest frame under the gauge transformations~\eqref{labeling:sp} and \eqref{labeling:tm}.
The pullback of $u^\alpha(\sigma)$ to the physical spacetime is written as
\begin{align}
	u^\mu(x) = \frac{\partial x^\mu}{\partial \sigma^\alpha} u^\alpha(\sigma),\label{def:umu} 
\end{align}
which is implicit for $\pi^\mu$.
In fact, 
\begin{align}
	\frac{\partial x^\mu }{ \partial \sigma^\alpha} = (K^{-1})^\mu{}_\alpha.
\end{align} 
To the leading order in the NG modes, we find
\begin{align}
	(K^{-1})^\mu{}_\alpha &= \delta^\mu{}_\alpha + \partial_\alpha \pi^\mu + \mathcal O(\partial \pi^2),
	\\
	b^{-1} & = \frac{1}{\sqrt{-g_{00}}}(1 - \partial_0\pi^0 + \mathcal O(\partial \pi^2)),
\end{align}
and thus, the fluid velocity in the physical spacetime is written as
\begin{align}
	u^0(x) = \frac{1}{\sqrt{-g_{00}}},~u^i(x) = \frac{\partial_0 \pi^i}{\sqrt{-g_{00}}} .\label{phys:vel:def} 
\end{align}
Note that we used $g_{0i}=\mathcal O(\partial \pi)$ for cosmological perturbations in an FLRW background spacetime.
To summarize, the internal gauge symmetry introduces a rest frame.
There, we constructed the fluid velocity $u^\mu$ expressed by the hydrodynamical mode $\pi^i$.

For a local thermal state, there is another scalar field, the local proper temperature field $\beta$, which depends on $x$ through $\sigma$.
Ref.~\cite{Crossley:2015evo} proposed to synchronize $\sigma^0$ to have
\begin{align}
	 \beta = \sqrt{-\bar \beta^2 h_{00}},\label{proptempgauge}
\end{align}
where $\bar \beta$ is the initial temperature introduced in Eq.~\eqref{def:rhobeta0}.
This condition is understood as a gauge condition to fix Eq.~\eqref{labeling:tm}, and the residual gauge symmetry is \eqref{labeling:sp} and 
\begin{align}
	\bar \sigma^0 = \sigma^0 + f(\sigma^a)
	\label{labeling:sptm}.
\end{align} 
In this gauge, the local temperature is written by the NG mode:
\begin{align}
	\beta(\sigma(x)) = b \bar \beta = \sqrt{-g_{00}}\bar \beta (1 + \partial_0\pi^0 + \mathcal O(\partial \pi^2) ).\label{def:loc:temp}
\end{align}
Thus, we introduced the local temperature associated with the fluid elements expressed by the remaining hydrodynamical variable $\pi^0$.

\subsection{Gradient expansion}\label{subsec:grad}

The effective Lagrangian can be constructed from $\pi^\mu$, $\pi_A^\mu$, $g_{\mu\nu}$, $g_{A\mu\nu}$, and their covariant derivatives.
$\pi_A^\mu$, $g_{\mu\nu}$, and $g_{A\mu\nu}$ are linear representations, while $\pi^\mu$ is nonlinear.
Hence, the covariant derivatives of $\pi^\mu$ are defined via $\beta$ and $u^\mu$.
From $\beta$, $u^\mu$, and $g_{\mu\nu}$, and their covariant derivatives, one finds the gradient expansion of an arbitrary classical energy-momentum tensor $T^{\mu\nu}$, and the classical effective action
\begin{align}
	 I_1[\pi,\pi_A;g,g_A] = \frac{1}{2}\int d^4x \sqrt{-\det g} T^{\mu\nu}G_{A\mu\nu}.\label{defI1}
\end{align}
To the lowest order in the gradient expansion, one can write  
\begin{align}
	T^{\mu\nu}_0 = \rho_0 u^\mu u^\nu + P_0\gamma^{\mu\nu},\label{perfect}
\end{align}
where $\rho_0$ and $P_0$ are functions of $\beta$ and we defined the induced metric
\begin{align}
	\gamma^{\mu\nu} \equiv g^{\mu\nu} + u^\mu u^\nu.
\end{align}
Dissipation is included in the next-to-leading order in the gradient expansion:
\begin{align}
\begin{split}
	T^{\mu \nu}_1 &= \left(\lambda_1 u^\rho \nabla_\rho \ln \beta  - \lambda_2 \nabla_\rho u^\rho \right)u^\mu u^\nu 
	\\
	&+ \left(\lambda_3 u^\rho \nabla_\rho \ln \beta  - \lambda_4\nabla_\rho u^\rho \right)\gamma^{\mu\nu}
	\\
	&
	+(\lambda_5\nabla_\sigma \ln \beta - \lambda_6  u^\rho\nabla_\rho  u_\sigma ) ( u^\mu\gamma^{\nu\sigma} + u^\nu\gamma^{\mu\sigma})
	\\
	&
	- \lambda_7 \gamma^{\mu\alpha}\gamma^{\nu\beta}\left(\nabla_\beta u_\alpha  + \nabla_\alpha u_\beta - \frac{2}{3}\gamma_{\alpha\beta}\nabla_\gamma u^\gamma \right),
\end{split}
	\label{emtt1grad}
\end{align}
where the anisotropic shear $\lambda_7$ describes dissipation.
$\lambda_{1}$ and $\lambda_{2}$ are the corrections to the energy density.
$\lambda_{3}$ and $\lambda_4$ are those to the pressure.
$\lambda_{5}$ and $\lambda_{6}$ are the momentum density.
In this way, we can expand the energy-momentum tensor to any order in the gradient expansion, and we find
\begin{align}
	T^{\mu\nu} = \sum_{n=0}^{\infty} T^{\mu\nu}_n.
\end{align}

\subsection{Local KMS symmetry}

So far, we have constructed the effective theory to the leading order in the advanced fields, which corresponds to classical hydrodynamics.
Dissipation is incorporated into $T^{\mu\nu}_{1}$ in this effective theory.
The stochastic noise is absent in classical hydrodynamics since the next-to-leading order in the advanced fields corresponds to the noise.
In the effective theory, to the leading order in the gradient expansion, $I_2$ is written as
\begin{align}
	 I_2[\pi,\pi_A;g,g_A] = \frac{i}{4}\int d^4x \sqrt{-\det g} W_0^{\mu\nu,\rho\sigma}G_{A\mu\nu}G_{A\rho\sigma}.\label{I2def}
\end{align}
$W_0^{\mu\nu,\rho\sigma}$ is lowest order in the derivatives.  
$I_2$ is invariant under the fluid spacetime diffeomorphism since $W_0^{\mu\nu,\rho\sigma}$ is constructed from $\beta$, $u^\mu$, $g_{\mu\nu}$. 
The diagonal spacetime diffeomorphism invariance of this action is manifest.
$I_2$ is invariant to $\mathcal O(\hbar^2)$ under the leading order gauge transformation generated by $\xi_A$.
By construction, $W_0^{\mu\nu,\rho\sigma} = W_0^{\nu\mu,\rho\sigma}= W_0^{\mu\nu,\sigma\rho}= W_0^{\nu\mu,\sigma\rho}$.
A general expression for $W^{\mu\nu,\rho\sigma}_0$ is 
\begin{align}
\begin{split}
		W_0^{\mu\nu,\rho\sigma} 
	=&c_1 u^\mu u^\nu u^\rho u^\sigma 
	+ c_2 ( \gamma^{\mu \nu}u^\rho u^\sigma + \gamma^{\rho\sigma}u^\mu u^\nu )
	\\
	&+ c_3 ( \gamma^{\mu \rho}u^\nu u^\sigma + \gamma^{\nu \rho}u^\mu u^\sigma +\gamma^{\mu \sigma}u^\nu u^\rho+\gamma^{\nu \sigma}u^\mu u^\rho)
	\\ 
	&+ c_4 \gamma^{\mu\nu}\gamma^{\rho\sigma}
	+ c_5\left( \gamma^{\mu\rho}\gamma^{\nu\sigma} 
	+\gamma^{\nu\rho}\gamma^{\mu\sigma} 
	-\frac{2}{3}\gamma^{\mu\nu}\gamma^{\rho\sigma}\right ),
\end{split}
\label{def:W0}
\end{align}
where $c_n$ are functions of $\beta$, which can be fixed by the local KMS symmetry and $I_1$.

Let us consider the consequences of the local KMS symmetry.
Following the proposal in Ref.~\cite{Crossley:2015evo}, we set the dynamical variables to zero, and fix the interaction for the local action.  
By definition, we find from the time reversal transformations for the metric:
\begin{align}
\begin{split}
	g_{00}(x^0,x^i) &\to g_{00}(-x^0,x^i),
	\\
	g_{0i}(x^0,x^i) &\to -g_{0i}(-x^0,x^i),
	\\
	g_{ij}(x^0,x^i) &\to g_{ij}(-x^0,x^i).	
\end{split}
\end{align}
The velocity in the absence of the dynamical variables is written as
\begin{align}
	u^\mu = \frac{\delta^{\mu}{}_0}{b},~u_\mu = \frac{g_{0\mu}}{b},~b=\sqrt{-g_{00}}
\end{align}
Then, the time reversal of the velocity is given as 
\begin{align}
\begin{split}
	u^0(x^0,x^i) &\to u^0(-x^0,x^i),
	\\
	u_0(x^0,x^i) &\to u_0(-x^0,x^i),
	\\
	u_i(x^0,x^i) &\to - u_i(-x^0,x^i),
\end{split}
\label{uideflow}
\end{align}
which yields the time reversal of the lowest-order energy-momentum tensor:
\begin{align}
\begin{split}
		T^{00}_0(x^0,x^i) &\to T^{00}_0(-x^0,x^i),
	\\
	T^{0i}_0(x^0,x^i) &\to -T^{0i}_0(-x^0,x^i),
	\\
	T^{ij}_0(x^0,x^i) &\to T^{ij}_0(-x^0,x^i).
\end{split}
\end{align}
Next, the induced metric transforms as
\begin{align}
\begin{split}
	\gamma^{00}(x^0,x^i) &\to \gamma^{00}(-x^0,x^i),
	\\
	\gamma^{0i}(x^0,x^i) &\to -\gamma^{0i}(-x^0,x^i),
	\\
	\gamma^{ij}(x^0,x^i) &\to \gamma^{ij}(-x^0,x^i).
\end{split}
\end{align}
Then, similarly, one can find the time reversal of $\nabla_\mu u_\nu$, which yields
\begin{align}
\begin{split}
		T^{00}_1(x^0,x^i) &\to -T^{00}_1(-x^0,x^i),
	\\
	T^{0i}_1(x^0,x^i) &\to T^{0i}_1(-x^0,x^i),
	\\
	T^{ij}_1(x^0,x^i) &\to -T^{ij}_1(-x^0,x^i).
\end{split}
\end{align}
$G_{A\mu\nu}=g_{A\mu\nu}$ transforms like $g_{\mu\nu}$. Hence, we find the time reversal of the local terms:
\begin{align}
	T_0^{\mu\nu}G_{A\mu\nu} &\to T_0^{\mu\nu}G_{A\mu\nu},
	\\
	T_1^{\mu\nu}G_{A\mu\nu} &\to - T_1^{\mu\nu}G_{A\mu\nu}.
\end{align}
If we assume the PT symmetry for the microscopic theory, one can find the same results more easily.
Next, we consider the imaginary time translation generated by $\xi_A^\mu =i\hbar \beta^\mu$ with $\beta^\mu \equiv \beta u^\mu$ in Eqs.~\eqref{xiAgmunu} and \eqref{xiAgmunuA}:
\begin{align}
	T^{\mu\nu} &\to T^{\mu\nu}+  \mathcal O(\hbar ^2),
	\\
	G_{A\mu\nu} &\to G_{A\mu\nu} + i\hbar \pounds_\beta g_{\mu\nu} + \mathcal O(\hbar ^3).
\end{align}
Combining the time reversal and the imaginary time translation, we obtain the following local KMS transformation:
\begin{align}
	T_0^{\mu\nu}G_{A\mu\nu} &\to T_0^{\mu\nu}G_{A\mu\nu} + i\hbar T_0^{\mu\nu} \pounds_\beta g_{\mu\nu} + \mathcal O(\hbar^3),
	\\
	T_1^{\mu\nu}G_{A\mu\nu} &\to - T_1^{\mu\nu}G_{A\mu\nu} - i\hbar T_1^{\mu\nu} \pounds_\beta g_{\mu\nu} + \mathcal O(\hbar^3),
	\\
\begin{split}
		W_0^{\mu\nu,\alpha \beta}G_{A\mu\nu}G_{A\alpha\beta} &\to W_0^{\mu\nu,\alpha \beta}G_{A\mu\nu}G_{A\alpha\beta} + 2i\hbar W_0^{\mu\nu,\alpha \beta}G_{A\mu\nu}\pounds_\beta g_{\alpha\beta}
	\\
	&  -\hbar^2 W_0^{\mu\nu,\alpha \beta}\pounds_\beta g_{\mu\nu}\pounds_\beta g_{\alpha\beta} + \mathcal O(\hbar^4).
\end{split}
\end{align}
The local KMS transformation of the effective Lagrangian is 
\begin{align}
\begin{split}
\delta_{\rm KMS} \mathcal L_{\rm eff}=& \frac{i\hbar}{2} T_0^{\mu\nu} \pounds_\beta g_{\mu\nu}
	-  \frac{i\hbar}{2} T_1^{\mu\nu} \pounds_\beta g_{\mu\nu} 
	\\
	&
	- \frac{i\hbar^2}{4} W_0^{\mu\nu,\alpha \beta}\pounds_\beta g_{\mu\nu}\pounds_\beta g_{\alpha\beta}
	\\
	&
	-  T_1^{\mu\nu}G_{A\mu\nu}
	- \frac{1}{2}\hbar W_0^{\mu\nu,\alpha \beta}G_{A\mu\nu}\pounds_\beta g_{\alpha\beta}
	+ \mathcal O(\hbar^3).	
\end{split}
\label{kmsloc}
\end{align}
The first-order advanced field terms are eliminated if
\begin{align}
	T_1^{\mu\nu}= - \frac{1}{2}\hbar W_0^{\mu\nu,\alpha \beta}\pounds_\beta g_{\alpha\beta},\label{fdrcomp}
\end{align}
which also cancels the $\mathcal O(\hbar^2)$ term in the first line.
The thermodynamical relation 
\begin{align}
	\rho_0 + P_0 = - \beta \frac{\partial P_0}{\partial \beta},\label{thermrel}
\end{align}
yields
\begin{align}
	 T_0^{\mu\nu} \nabla_\mu \beta_{\nu} = \nabla_\mu (P_0\beta^\mu),
\end{align}
which implies, in other words, the local KMS symmetry imposes Eq.~\eqref{thermrel}.
Then, the local KMS transformation of the effective Lagrangian turns into a total derivative 
\begin{align}
	\delta_{\rm KMS} \mathcal L_{\rm eff}=&  \nabla_\mu (P_0\beta^\mu) + \mathcal O(\hbar^3).
\end{align}
The fluctuation-dissipation relation~\eqref{fdrcomp} fixes $W_0$ as follows.
We have
\begin{align}
	\pounds_\beta g_{\mu\nu} = \nabla_\mu \beta_\nu + \nabla_\nu \beta_\mu.
\end{align}
Then, one finds
\begin{align}
\begin{split}
	u^\mu u^\nu u^\rho u^\sigma \nabla_\sigma \beta_\rho &= -u^\mu u^\nu u^\rho\nabla_\rho \beta,
	\\
	( \gamma^{\mu \nu}u^\rho u^\sigma + \gamma^{\rho\sigma}u^\mu u^\nu )\nabla_\sigma \beta_\rho &= -\gamma^{\mu\nu} u^\sigma \nabla_\sigma \beta + \beta u^\mu u^\nu \nabla_\rho u^\rho,
	\\
	( \gamma^{\mu \rho}u^\nu u^\sigma + \gamma^{\nu \rho}u^\mu u^\sigma ) \nabla_\sigma \beta_\rho & = \beta u^\sigma \nabla_\sigma(u^\mu u^\nu) ,
	\\
	(\gamma^{\mu \sigma}u^\nu u^\rho+\gamma^{\nu \sigma}u^\mu u^\rho)\nabla_\sigma \beta_\rho & =-( u^\mu\gamma^{\nu\sigma} + u^\nu\gamma^{\mu\sigma})\nabla_\sigma \beta,
	\\
	\gamma^{\mu\nu}\gamma^{\rho\sigma}\nabla_\sigma \beta_\rho & =\beta   \gamma^{\mu\nu} \nabla_\rho u^\rho,
	\\
	( \gamma^{\mu\rho}\gamma^{\nu\sigma} 
	+\gamma^{\nu\rho}\gamma^{\mu\sigma} )\nabla_\sigma \beta_\rho & = \beta \gamma^{\mu\rho}\gamma^{\nu\sigma}(\nabla_\sigma u_\rho + \nabla_\rho u_\sigma).
\end{split}
\end{align}
These projections and Eq.~\eqref{fdrcomp} yield
\begin{align}
\begin{split}
	 T_1^{\mu\nu} &= u^\mu u^\nu\left(c_1\hbar \beta  u^\rho\nabla_\rho \ln \beta - c_2\hbar \beta  \nabla_\rho u^\rho \right)
	 \\
	 &+\gamma^{\mu\nu}(c_2 \hbar \beta u^\sigma \nabla_\sigma \ln \beta  - c_4\hbar\beta    \nabla_\rho u^\rho)
	 \\
	 &+( u^\mu\gamma^{\nu\sigma} + u^\nu\gamma^{\mu\sigma}) (c_3 \hbar \beta \nabla_\sigma \ln \beta - c_3\hbar \beta  u^\rho\nabla_\rho  u_\sigma  )	
	 \\
	 & - \gamma^{\mu\rho}\gamma^{\nu\sigma}\left(\nabla_\sigma u_\rho + \nabla_\rho u_\sigma-\frac{2}{3}\gamma_{\rho\sigma}\nabla_\lambda u^\lambda\right)c_5\hbar \beta.
\end{split}
\end{align}
Comparing this expression with Eq.~\eqref{emtt1grad}, we find the fluctuation-dissipation relations~\cite{Crossley:2015evo}:
\begin{align}
\begin{split}
	c_1  & =  \frac{\lambda_1}{\hbar \beta},~
	c_2  = \frac{\lambda_2}{\hbar\beta} = \frac{\lambda_3}{\hbar \beta},
	\\
	c_3 & = \frac{\lambda_5}{\hbar \beta}=\frac{\lambda_6}{\hbar \beta},
	~
	c_4  = \frac{\lambda_4}{\hbar \beta},
	~
	c_5  = \frac{\lambda_7}{\hbar \beta}. 
\end{split}
	\label{fdrc}
\end{align}
These relations will be used to identify the noise in the CMB in the next section.
Interestingly, the KMS condition yields the classical contribution $I_2 =\mathcal O(\hbar)$ to the second order in the $A$ fields.
We got the additional constraints $\lambda_2 = \lambda_3$ known as the non-linear Onsager relations, and $\lambda_5=\lambda_6$.
In the following sections, we introduce the following parametrizations:
\begin{align}
	\lambda_\rho \equiv \lambda_1 -3 \lambda_2,~\lambda_P \equiv \lambda_2 - 3\lambda_4,~\lambda_4 \equiv \zeta,~\lambda_5\equiv \kappa~\lambda_7\equiv \eta.\label{newparams}
\end{align}

\subsection{Langevin equation}

At the semiclassical order, the Langevin equation is useful to describe the noise.
One can introduce a Gaussian distribution 
\begin{align}
	\mathcal P[\xi]\equiv Z_A^{-1}e^{-\frac{1}{4}\int d^4 x \sqrt{-g}\xi^{\mu\nu} (W_0^{-1})_{\mu\nu,\rho\sigma}\xi^{\rho\sigma}},~ Z_A \equiv \int D\xi e^{-\frac{1}{4}\int d^4 x \sqrt{-g}\xi^{\mu\nu} (W_0^{-1})_{\mu\nu,\rho\sigma}\xi^{\rho\sigma}},
	\label{xidist}
\end{align}
where we defined the inverse of $W_0$ as 
\begin{align}
	W_0^{\mu\nu,\alpha \beta}(W_0^{-1})_{\alpha \beta,\rho\sigma} = (W_0^{-1})_{\rho\sigma,\alpha \beta} W_0^{\alpha \beta, \mu\nu} = \frac{1}{2}( \delta^\alpha{}_\rho \delta^\beta{}_\sigma + \delta^\alpha{}_\sigma \delta^\beta{}_\rho). 
\end{align}
Then, one can write the original path integral \eqref{partialpathint} as 
\begin{align}
\begin{split}
		e^{W[g,g_A]} &= Z_A^{-1} \int D\pi_A D\pi D\xi 
	\\
	&\times e^{i \int d^4 x \sqrt{-g}\left[ \frac{1}{2} T^{\mu\nu}G_{A\mu\nu} + \frac{i}{4}G_{A\mu\nu} W^{\mu\nu,\rho\sigma}G_{A\rho \sigma}+ \frac{i}{4} \xi^{\mu\nu} (W_0^{-1})_{\mu\nu,\rho\sigma}\xi^{\rho\sigma} \right]}.
\end{split}
\end{align}
Redefining $\xi$ as 
\begin{align}
	\xi^{\mu \nu} \to \xi^{\mu \nu} - i W^{\mu\nu,\rho\sigma}G_{A\rho \sigma },
\end{align}
one finds
%\begin{align}
%	e^{W[g,g_A]} &= \int D\pi_A D\pi D\xi \mathcal P[\xi]e^{i \int d^4 x\sqrt{-g} \frac{1}{2}(T^{\mu\nu} + \xi^{\mu\nu})G_{A\mu\nu}}.
%\end{align}
%Integrating by parts, we obtain
\begin{align}
	e^{W[g,g_A]} &= \int D\pi_A D\pi D\xi \mathcal P[\xi]e^{i \int d^4 x\sqrt{-g} \frac{1}{2}(T^{\mu\nu} + \xi^{\mu\nu})g_{A \mu\nu}}e^{-i \int d^4 x\sqrt{-g} \nabla_\mu (T^{\mu\nu} + \xi^{\mu\nu}) \pi_{A\nu}}.\label{pathintxi}
\end{align}
Introducing the delta functional, we get
\begin{align}
	e^{W}	&= \int  D\pi D\xi \mathcal P[\xi]e^{i \int d^4 x\sqrt{-g} \frac{1}{2}(T^{\mu\nu} + \xi^{\mu\nu})g_{A \mu\nu}}\delta[ \nabla_\mu (T^{\mu\nu} + \xi^{\mu\nu}) ].
\end{align}
Then, we find the stochastic equation of motion
\begin{align}
	\nabla_\mu (T^{\mu\nu} + \xi^{\mu\nu})=0,\label{eomxi}
\end{align}
with the probability distribution \eqref{xidist}.
Notice that we ignored the Jacobian from $\xi^{\mu\nu}$ to the equation of motion.
We may ignore the Jacobian at the tree level order as discussed in Ref.~\cite{Crossley:2015evo}.
The partition function of $\xi$ is found by setting $T^{\mu\nu}\to 0$ and $D\pi_AD\pi \to 1$ in Eq.~\eqref{pathintxi}:
\begin{align}
	e^{W_\xi[g_A]} \equiv  \int  D\xi \mathcal P[\xi]e^{\frac{i}{2} \int d^4 x\sqrt{-g} \xi^{\mu\nu}g_{A\mu\nu}}.
\end{align}
Then, correlation functions of $\xi$ are obtained as
\begin{align}
	\langle \xi^{\mu\nu}(x)\cdots \rangle = \left[\frac{2}{\sqrt{-g(x)}}\frac{\delta}{i\delta g_{A\mu\nu}(x)}\right]\cdots W_\xi[g_A].
\end{align}
By construction, one immediately finds 
\begin{align}
	W_\xi[g_A] =  -\frac{1}{4}\int d^4 x \sqrt{-g} g_{A\mu\nu} W^{\mu\nu,\rho\sigma}g_{A\rho \sigma} ,
\end{align}
hence we obtain
\begin{align}
	\langle \xi^{\mu\nu}(x)\xi^{\rho\sigma}(y)\rangle  =  W^{\mu\nu,\rho\sigma}(x)\delta^{(4)}(x-y). \label{2ptxi}
\end{align}
Now, we have obtained the stochastic equation of motion for the hydrodynamical variable $\pi^\mu$, and the noise average.
This is the end of the preparation for the cosmological application.

\section{Effective theory for cosmological fluids}
\label{sec:cosmofluid}

As discussed in the previous section, the effective field theory of fluctuating hydrodynamics is based on the Schwinger-Keldysh formalism.
In this formalism, we compute the expectation values of observables; therefore, the path integral is defined for the CPT.
The CPT doubles the dynamical variables as well as the diffeomorphism structure in the original UV action.
In the Keldysh basis, the doubled structure is separated into diagonal and nondiagonal components.
The hydrodynamical modes are the NG modes associated with the diagonal and nondiagonal spacetime diffeomorphisms generated by Eqs.~\eqref{def:xiR} and~\eqref{def:xiA}.
These NG modes write the hydrodynamical effective action, which describes the dynamics of sound waves and the stochastic noise.

The EFT parameters in the hydrodynamical action cannot be determined in the effective theory itself.
Hence, we need input from the UV theory or experiments.
In the present case, the Boltzmann theory of cosmological fluids is assumed.
For a given classical energy-momentum tensor $T^{\mu\nu}$ in the kinetic theory, one can read the dissipative coefficients~\eqref{newparams}.
Then, using the fluctuation dissipation relation~\eqref{fdrc}, one finds the noise $W_0^{\mu\nu,\rho\sigma}$.
The stochastic noise arises in the stochastic equation of motion~\eqref{eomxi}, and the 2-point correlation function of the noise is written as Eq.~\eqref{2ptxi}.
If $W_0^{\mu\nu,\rho\sigma}$ depends only on time, the Fourier transformation of Eq.~\eqref{2ptxi} yields Eq.~\eqref{2ptanticipate}.

Derivation in the previous section applies to a general curved spacetime, and hence, it straightforwardly applies to cosmological perturbation theory.
In this section, we consider the physical spacetime metric perturbed around an FLRW spacetime.
We will see the fluctuation-dissipation relation in the expanding universe.

\subsection{Setup}
For a cosmological background, we consider a physical spacetime metric perturbed around a flat FLRW spacetime.
A choice of a flat FLRW spacetime is not unique due to the physical spacetime diffeomorphism invariance.
Here we write the metric in the conformal Newtonian gauge~\cite{Ma:1995ey}:
\begin{align}
	g_{00} = -a^2 (1 + 2\phi),~g_{0i}=g_{i0}=0,~g_{ij} = a^2 (1 + 2\psi)\delta_{ij}.\label{confgauge}
\end{align}
Note that the gauge freedom in cosmological perturbations is distinct from the gauge freedom in the fluid spacetime.
We only consider the scalar perturbations for simplicity.
The NG modes are perturbations around the homogeneous and isotropic fluid, so we regard them as first-order perturbations.

Hereafter, expansion in cosmological perturbations is truncated at first order.
In the conformal Newtonian gauge, the local proper temperature~\eqref{def:loc:temp} is written as
\begin{align}
	\beta = a\bar \beta(1  +\phi+\partial_0\pi^0),
\end{align}
where $\pi^\mu$ is the NG mode associated with the diagonal physical spacetime diffeomorphism generated by Eq.~\eqref{def:xiR}. 
The fluid velocity~\eqref{phys:vel:def} is expanded into
\begin{align}
\begin{split}
		u^0(x) &= a^{-1}(1-\phi),~u^i(x) = a^{-1} \partial_0 \pi^i, 
\\
	u_0(x) &= - a(1+\phi),~u_i(x) = a\partial_0 \pi_i. 
\end{split}
\end{align}
The spatial indexes for cosmological perturbations are raised and lowered by the flat spatial metric $\delta^{ij}$ and $\delta_{ij}$.
For example, $\pi^i = \delta^{ij} \pi_j$ and $\partial^i = \delta^{ij}\partial_j$.
This index notation is common in cosmological perturbation theory, but it should be distinct from the notation for the spacetime tensors in the previous sections.
The induced metric associated with the fluid velocity is
\begin{align}
	\gamma^{00}= 0,~\gamma^{0i} = \gamma^{i0} = a^{-2}\partial_0 \pi^i,~\gamma^{ij} = a^{-2}(1- 2\psi)\delta^{ij}.
\end{align}
The projection tensor to the spatial hypersurface perpendicular to the fluid 4-velocity is
\begin{align}
	\gamma^0{}_0 = 0,~\gamma^0{}_i=\partial_0\pi_i,~ \gamma^i{}_0= -\partial_0\pi^i,~\gamma^i{}_j =\delta^i{}_j.
\end{align}
The Christoffel symbol 
\begin{align}
\Gamma^\mu{}_{\nu\rho}\equiv \frac12g^{\mu\alpha}\left(\partial_\rho g_{\alpha\nu}+\partial_\nu g_{\alpha\rho}-\partial_\alpha g_{\nu\rho} \right).
\end{align}
is given as
\begin{align}
\begin{split}
\Gamma^0{}_{00}=&\mathcal H+\partial_0 \phi,~\Gamma^0{}_{0i}=\partial_i\phi,~
\Gamma^0{}_{ij}=
(
\mathcal H-2\mathcal H\phi+2\mathcal H\psi+\partial_0\psi)\delta_{ij},\\
\Gamma^{i}{}_{00}=&
\partial^i\phi,~\Gamma^i{}_{0j}=(\mathcal H+\partial_0\psi)\delta^i{}_{j},
~\Gamma^i{}_{jk}=-\partial^i\psi\delta_{jk}+\partial_k\psi\delta^i{}_{j}+\partial_j\psi\delta^i{}_{k},	
\end{split}
\end{align}
with the conformal Hubble parameter $\mathcal H \equiv \partial_0 a/a$.
In the following calculation, we often use 
\begin{align}
	\nabla_\mu u_{\nu} = \partial_\mu u_\nu - \Gamma^0_{\mu\nu}u_0- \Gamma^k_{\mu\nu}u_k,
\end{align}
which is evaluated as
\begin{align}
\begin{split}
		\nabla_{0}u_0 &= 0 ,~\nabla_{0}u_i =a \partial_0^2\pi_i + a\partial_i\phi,~
	\nabla_{i}u_0 =-a \mathcal H \partial_0 \pi_i,
	\\	 
	\nabla_{i}u_j &=a \delta_{ij}
\mathcal H  + a \partial_0 \partial_i \pi_j + a \delta_{ij}(
-  \mathcal H \phi + 2\mathcal H \psi +\partial_0\psi) .
\end{split}
\end{align}
The covariant divergence of $u^\mu$ is 
\begin{align}
	\nabla_\rho u^\rho 
	&=3a^{-1}\left(\mathcal H -\mathcal H\phi +\partial_0\psi  +\frac{1}{3}\partial_i \partial_0\pi^i\right),
\end{align}
and the directional derivative of the local temperature to the fluid velocity is
\begin{align}
	u^\rho \nabla_\rho \ln \beta 
	&=  a^{-1} (\mathcal H - \mathcal H \phi+ \partial_0 \phi + \partial_0^2 \pi^0). 
\end{align}
A function of $\beta$ is expanded into 
\begin{align}
	f(\beta) = f(a\bar \beta) + (\phi+\partial_0\pi^0) f'(a\bar \beta),~' = \frac{d}{d \ln a}.
\end{align}

\subsection{Matching EFT coefficients}
\label{subsec:matching}
Using these equations Eqs.~\eqref{emtt1grad}, and ~\eqref{newparams}, one finds the energy-momentum tensor order by order in cosmological perturbations.
At zeroth order, one finds $T^{0i}_{(0)}=T^{i0}_{(0)}=0$, and
\begin{align}
\begin{split}
	T^{00}_{(0)} &=a^{-2}\rho,~\rho\equiv \rho_0 +a^{-1}\mathcal H  \lambda_\rho	  \label{Edef},
\end{split}
\\
\begin{split}
	T^{ij}_{(0)} &=a^{-2}P,~P\equiv \left[ P_0 +  a^{-1}\mathcal H \lambda_P  \right]\delta^{ij}\label{Pdef},
\end{split}
\end{align}
where a lower index in a bracket implies the order in cosmological perturbations.
Thus, some first-order gradient terms may contribute to the zeroth order of cosmological perturbation.
By construction in Section~\ref{subsec:grad}, $\rho_0$, $P_0$, $\lambda_\rho$ and $\lambda_P$ are functions of $a\bar \beta$, and they depend on time through $a$.
It can be shown that $\rho$ and $P$ are written as functions of $a$ for equations of states independent from time derivatives of $a$.
One may regard $\lambda_\rho$ and $\lambda_P$ are the time scales of matter variation.
$\lambda_\rho$ and $\lambda_P$ will be present during Big Bang nucleosynthesis or recombination as the matter contents vary.
However, these events are slow, compared to the Hubble time.
For more drastic events, such as phase transition or electron-positron pair annihilation, we may have $\mathcal H \lambda_\rho\gg 1$ and $\mathcal H \lambda_P\gg 1$. 
Given a set of background evolution in the Boltzmann theory, one can find $\rho_0$, $P_0$, $\lambda_\rho$ and $\lambda_P$ by fitting the solutions, assuming the time dependence in Eqs.~\eqref{Edef} and \eqref{Pdef}.

Next, we identify the NG mode and fit $\kappa$, $\zeta$ and $\eta$ by computing the first order energy momentum tensor:
\begin{align}
\begin{split}
	T^{00}_{(1)} &= a^{-2}(\rho_0' +a^{-1}\mathcal H \lambda_\rho') (\phi+\partial_0\pi^0)       + a^{-2} \rho_0  \phi  
	\\
	&   
	 + a^{-3} (\lambda_\rho+3\lambda_P + 9\zeta) (\partial_0 \phi + \partial_0^2 \pi^0) 
	\\
	& - a^{-3}(\lambda_P+3 \zeta) (3 \partial_0\psi + \partial_i \partial_0\pi^i)
	,
\end{split}
	\label{T00}
\\
\begin{split}
	T^{0i}_{(1)} &= a^{-2}\left(\rho_0 + P_0 + a^{-1}\mathcal H\lambda_\rho + a^{-1}\mathcal H \lambda_P\right)\partial_0 \pi^i  
	\\
	&+a^{-3} \kappa  
	 (\partial_0\partial^i\pi^0  -  \partial_0^2 \pi^i )  ,
	 	\end{split}
\label{T0i}  
\\
\begin{split}
	T^{ij}_{(1)} &= a^{-2} \left[ P_0' +  a^{-1}\mathcal H \lambda_P'  \right](\phi+\partial_0\pi^0) \delta^{ij} +  a^{-2}P_0 \phi  \delta^{ij}
	\\
	&
	+ a^{-3}(\lambda_P +3\zeta) (\partial_0 \phi + \partial_0^2 \pi^0) \delta^{ij} 
	  \\
	  &
	  -  a^{-3} \zeta \left(3\partial_0\psi  +\partial_k \partial_0\pi^k\right) \delta^{ij}
	\\
	&
	- \eta a^{-3}\left(  \partial^i \partial_0  \pi^j+  \partial^j \partial_0  \pi^i 	  - \frac{2}{3}\delta^{ij} \partial_k \partial_0\pi^k \right).
\end{split}
	\label{Tij}
\end{align}
As discussed above, $\rho_0$, $P_0$, $\lambda_\rho$ and $\lambda_P$ in Eqs.~\eqref{T00} to \eqref{Tij} are are already determined at zeroth order.
$\phi$ and $\psi$ are determined by the Boltzmann theory.
For simplicity, let us consider Eqs.~\eqref{T00} to \eqref{Tij} in Fourier space.
As we only consider scalar perturbations, one finds four independent equations for each time step and Fourier mode: Eqs.~\eqref{T00}, the spatial divergence of \eqref{T0i}, the spatial trace of \eqref{Tij} and traceless part of \eqref{Tij}.
In addition to the matching conditions~\eqref{T00} to \eqref{Tij}, the linearized continuity and Euler equations are used to eliminate the time derivatives of $\partial_0\pi^0$, $\partial_i\partial_0\pi^i$.
Hence, there are five independent variables: $\kappa$, $\zeta$, $\eta$, $\partial_0\pi^0$, and $\partial_i\partial_0\pi^i$, so we cannot totally fix the EFT parameters in general.
It was found that $\kappa\neq 0$ is incompatible when explaining hydrodynamics by course-graining the microscopic Boltzmann theory~\cite{Tsumura:2006hnr}.
In the old days, it was also found that $\kappa \neq 0$ shows instability in the hydrodynamical perturbations~\cite{Hiscock:1985zz}.
Therefore, we set $\kappa=0$.
In this case, one can fit all EFT parameters in the linearized equations.

\subsection{Equation of motion with noise}

Given a set of the EFT parameters, we are able to discuss the noise.
By integrating Eq.~\eqref{defI1} by parts, the classical part of the effective action is written as
\begin{align}
	I_1 = \int d^4 x\sqrt{-\det g} (\nabla_\mu T^{\mu \nu}) \pi_{A\nu},
\end{align}
where we set $g_{A\mu\nu}=0$, for simplicity.
Then, one can find the effective action order by order in cosmological perturbations.
Notice that the reference index position for the advanced NG mode is $\pi_{A\mu}$.
Hence, hereafter, we use $\pi_A^i \equiv \delta^{ij}\pi_{Aj} $ in the context of cosmological perturbations. 

The zeroth order action is absent since the Keldysh action starts from the first order in the advanced NG mode due to the unitarity condition~\eqref{Wgg=0}.
At first order in cosmological perturbations, we find
\begin{align}
	 I_{1(1)} = \int d^4x  a^2 \pi_{A0}\left[ \partial_0 \rho + 3\mathcal H(\rho+P)  \right], 
\end{align}
The background equation of motion, $\delta I_{1(1)}/\delta \pi_{A0}=0$, is found as
\begin{align}
	\partial_0 \rho + 3\mathcal H(\rho+P) = 0.
\end{align}
This is the standard continuity equation in the FLRW spacetime.

The second-order action is separated into four parts:
\begin{align}
	I_{(2)} = I_{1(2)\rm T} + I_{1(2)\rm S} + I_{2(2)\rm T}+ I_{2(2)\rm S}.
\end{align}
The time parts are written as
\begin{align}
	I_{1(2)\rm T} &= \int d^4 x a^2  \rho \left( \partial_0 \delta_\rho + 3( 1 + w) \partial_0\psi + 3 \mathcal H  w (\delta_P-\delta_\rho)  + (1+w)\theta
		\right)  \pi_{A0},
		\\
			I_{2(2)\rm T} &= \int d^4 x a^2  \rho \left( \partial_0 \xi_\rho  + 3 \mathcal H  w (\xi_P-\xi_\rho)  + (1+w)\xi_\theta
		\right)  \pi_{A0},
\end{align}
where we write $\delta_\rho$, $\delta_P$ and $\theta$ by the NG modes through  
\begin{align}
\begin{split}
		\rho \delta_\rho & = (\rho_0' +a^{-1}\mathcal H \lambda_\rho') (\phi+\partial_0\pi^0)       - a^{-1} \mathcal H \lambda_\rho  \phi  
	\\
	&   
	 + a^{-1} (\lambda_\rho+3\lambda_P + 9\zeta) (\partial_0 \phi + \partial_0^2 \pi^0) 
	\\
	& - a^{-1}(\lambda_P+3 \zeta) (3 \partial_0\psi + \partial_i \partial_0\pi^i),	
\end{split}
	\\
	\begin{split}
	P\delta_P & = 
	\left[ P_0' +  a^{-1}\mathcal H \lambda_P'  \right](\phi+\partial_0\pi^0) \delta^{i}{}_j -  a^{-1}\mathcal H \phi \lambda_P 
	\\
	&
	+ a^{-1}(\lambda_P +3\zeta) (\partial_0 \phi + \partial_0^2 \pi^0) 
	  \\
	  &
	  -  a^{-1} \zeta \left(3\partial_0\psi  +\partial_k \partial_0\pi^k\right) ,		
	\end{split}
	\\
	\begin{split}
			(\rho+P)\theta & =  \left(\rho_0 + P_0 + a^{-1}\mathcal H\lambda_\rho + a^{-1}\mathcal H \lambda_P\right)\partial_0 \partial_i\pi^i  
	\\
	&+a^{-1} \kappa  
	 (\partial^2\partial_0\pi^0  -  \partial_0^2 \partial_i\pi^i ) ,
	\end{split}
\end{align}
and we write $\xi_\rho$, $\xi_P$ and $\xi_\theta$ as  
\begin{align}
		\rho\xi_\rho & = a^{2}\xi^{00},
	\\
	P\xi_P & = \frac{1}{3} a^{2}\delta_{ij}\xi^{ij},
	\\
	(\rho+P)\xi_\theta & = a^2 \partial_i \xi^{i0}.
\end{align}
In the above expressions, we reduce the number of terms by eliminating the background equation of motion.
With $\pi_{Ai} = \partial_i \pi_A$, the spatial parts are recast into 
\begin{align}
	I_{1(2)\rm S} &=\int d^4 x a^2  (\rho + P) \left ( \partial_0 \theta +\mathcal H \theta + \frac{w}{1+w}\left( \partial^2 \delta_P + \theta \partial_0 \ln P\right)- \partial^2 \Sigma  + \partial^2 \phi  \right)  \pi_{A},
	\\
		I_{2(2)\rm S} &=\int d^4 x a^2  (\rho + P) \left ( \partial_0 \xi_\theta +\mathcal H \xi_\theta + \frac{w}{1+w}\left( \partial^2 \xi_P + \xi_\theta \partial_0 \ln P\right)- \partial^2 \xi_\Sigma    \right)  \pi_{A},
\end{align}
where we write $\Sigma$ and $\xi_\Sigma$ as
\begin{align}
	(\rho+P) \Sigma & = \frac{4}{3}\eta a^{-1} \partial_i \partial_0 \pi^i,\label{defSigma}
	\\ 
	(\rho+P) \xi_\Sigma & = - \partial^{-2}\left(\partial_i\partial_j - \frac{1}{3}\partial^2\delta_{ij}\right)a^2 \xi^{ij}. 
\end{align}
Then, the second-order action yields the first-order continuity and Euler equations in the presence of the stochastic noise:
\begin{align}
		\partial_0 \hat \delta_\rho + 3( 1 + w) \partial_0\psi + 3 \mathcal H  w (\hat \delta_P-\hat \delta_\rho)  + (1+w)\hat \theta &= 0,\label{mb:cont}
		\\
	\partial_0 \hat \theta +\mathcal H \hat \theta + \frac{w}{1+w}\left( \partial^2 \hat \delta_P + \hat \theta \partial_0 \ln P\right)- \partial^2 \hat \Sigma  + \partial^2 \phi& = 0,\label{mb:ue}
\end{align}
where a hat implies
\begin{align}
	\hat X \equiv X + \xi_X.
\end{align}
Setting the noise to zero, Eqs.\eqref{mb:cont} and \eqref{mb:ue} are consistent with the linearized continuity and Euler equations presented in Ref.~\cite{Ma:1995ey}.
From Eqs.~\eqref{def:W0}, \eqref{fdrc}, and \eqref{2ptxi}, one finds the following correlation functions
\begin{align}
	\langle \xi_\rho \xi_\rho\rangle'	& =\frac{\lambda_1}{\hbar \beta \rho^2} ,~\langle \xi_\rho \xi_P\rangle'	=\frac{\lambda_3}{\hbar \beta \rho P},
	~\langle \xi_\rho \xi_\theta \rangle'	=0 ,\label{xiExiE}
		~
	\langle \xi_\rho \xi_\Sigma \rangle'	 =0 ,
\\
	\langle \xi_P \xi_P\rangle'	& =\frac{\zeta}{\hbar \beta P^2} ,
	~
	\langle \xi_P \xi_\theta \rangle'	 =0 ,
		~
	\langle \xi_P \xi_\Sigma \rangle'	=0 \label{xiPxiP},
\\
	\langle \xi_\theta \xi_\theta \rangle'	&=\frac{\kappa \partial^2}{\hbar \beta (\rho+P)^2} ,\label{144}
~
	\langle \xi_P \xi_\Sigma \rangle'	 =0 ,
\\
		\langle \xi_\Sigma \xi_\Sigma \rangle'	&=\frac{4\eta }{3\hbar \beta (\rho+P)^2}\label{xisigxisig} ,
\end{align}
where a prime on a bracket implies we omitted the 4-dimensional delta function.
The derivative operator in Eq.~\eqref{144} is multiplied by the spatial delta function, which reduces to $k^2$ in Fourier space.
One can solve Eqs.~\eqref{mb:cont} and \eqref{mb:ue} with the linearized Einstein equations in the presence of the stochastic noise on the right-hand side.
The correlation functions are evaluated using Eqs.~\eqref{xiExiE} to \eqref{xisigxisig}.
In fact, solving a Boltzmann-Einstein system is a hard task.
In most cases, we need numerical calculations, and thus, the EFT coefficients are found numerically.
Therefore, further details can be discussed for specific models.

\section{Photon baryon plasma}

\label{pbp}

Let us consider photon-baryon plasma, which is the major contribution of the cosmological fluid during/before recombination.
The kinetic theory for this plasma is established in cosmology~\cite{Dodelson:2003ft}.
Baryons are subdominant components in the background evolution well before recombination.
In this case, many EFT coefficients are simplified.
At zeroth order in cosmological perturbations, we find
\begin{align}
	\rho_0 = \frac{g_*\pi^2}{30a^4 \bar \beta^4},~P_0=\frac{\rho_0}{3},\lambda_\rho=0,\lambda_P=0,\label{def:thdens}
\end{align}
where $g_*$ is a number of relativistic species.
On the other hand, the dimensionless fluctuations of baryons, such as the over-density or velocity, are not necessarily subdominant.
Photons and baryons are tightly coupled electromagnetically, and hence, these fluids move together.
Therefore, $\delta_b\approx \delta_\gamma$ and $\theta_b \approx \theta_\gamma$ are realized.
However, the baryon contributions to the total $\delta_\rho$ and $\theta$ are suppressed by $R\to 0$ defined in Eq.~\eqref{defR}.
Hence, the temperature fluctuation writes the density fluctuation:
\begin{align}
	\delta_\rho &= -4(\phi+ \partial_0 \pi^0).
\end{align}
Baryons are nonrelativistic, and hence, the pressure is zero.
Therefore, the total pressure is given by the thermal photon pressure:
\begin{align}
	\delta_P = \delta_\rho,~\zeta = 0.
\end{align}
As discussed at the end of Section~\ref{subsec:matching}, we set $\kappa=0$.
In this case, regardless of the other EFT coefficients, one can write
\begin{align}
		\theta & = \partial_i\partial_0\pi^i.
\end{align}
In the presence of radiative anisotropic pressure $\eta\neq 0$, Eqs.~\eqref{Tij} and \eqref{defSigma} yield
\begin{align}
	\Sigma = \frac{\eta \theta}{a\rho}.\label{sigmaeta}
\end{align}
Notice that the second unitarity condition~\eqref{uni2} implies $\eta>0$.

We focus on the sub-horizon dynamics of the hydrodynamical modes, so we set $\phi=\psi=0$.
We can solve Eqs.~\eqref{mb:cont} for $\partial_0\pi^0$ as
\begin{align}
			 \partial_0\pi^0    = \frac{1}{3} \partial_i  \pi^i ,\label{mb:cont2}
\end{align}
Then Eq.~\eqref{mb:ue} is recast into 
\begin{align}
	\partial_0^2  \partial_i  \pi^i  - \frac{1}{3} \partial^2  \partial_i  \pi^i   - \frac{\eta \partial^2}{a\rho} \partial_0 \partial_i\pi^i  = \partial^2 \xi_\Sigma.\label{mb:ue2}
\end{align}
%$a^{-1}\partial_0$ in the third term implies dissipation happens in the physical time $dt = adx^0$, not in the conformal time $x^0$.
Let us introduce $\partial^i \Pi \equiv \pi^i$ and the Hubble parameter $H\equiv a^{-1}da/dt$ with the physical time $dt = adx^0$, the equation of motion in Fourier space is obtained as
\begin{align}
	\partial_t^2 \Pi_{\mathbf k}  + \frac{k^2/a^2}{3} \Pi_{\mathbf k}   + (H+\gamma_{k/a}) \partial_t \Pi_{\mathbf k}   = a^{-2}\xi_{\Sigma,\mathbf k},\label{mb:ue3}
\end{align}
where we introduced 
\begin{align}
	\gamma_{k/a} \equiv \frac{\eta k^2}{\rho a^2}.
\end{align}
In the absence of $\gamma_k$ and $\xi_\Sigma$, the dispersion relation reduces to $\omega^2 = k^2/3$; we find sound wave with the sound velocity $1/\sqrt{3}$.
For sound waves with $H \ll \gamma_{k/a}$, one can ignore the cosmic expansion.
In this case, one may ignore the time dependence of the coefficients since they depend on time through $a(t)$.
Hence, we set $a=1$.
Then the retarded Green function is found as
\begin{align}
	G_k(t) = \frac{e^{-\frac{\gamma_k t}{2}} \sin\left(t\sqrt{k^2/3-\gamma^2_k/4}\right)\Theta(t)}{\sqrt{k^2/3 -\gamma^2_k/4}},
\end{align}
where $\Theta$ is the Heaviside step function.
Notice that $k$ is now physical momentum.
The inhomogeneous solution is integrated to
\begin{align}
	\Pi_{\mathbf k}(t) = \int_0^{t} d\bar t G_k(t-\bar t) \xi_{\Sigma ,{\mathbf k}}(\bar t).
\end{align}
We drop the homogenous solution, which corresponds to the primordial perturbation.
Eq.~\eqref{xisigxisig} yields
\begin{align}
	\langle \xi_{\Sigma, {\mathbf k}}(t) \xi_{\Sigma, \bar {\mathbf k}} (\bar t) \rangle	&=(2\pi)^3\delta^{(3)}(\mathbf k+ \bar{\mathbf k})\delta(t - \bar t) \frac{3\eta}{4 \beta \rho_0^2}. 	\label{xixidede}
\end{align}
Combining Eqs.~\eqref{xixidede} and 
\begin{align}
	\lim_{t\gamma_k \to \infty}\int^{t}_0 d\bar t^0G_k(t -\bar t)^2 =\frac{3\rho_0}{2 k^4 \eta}, 
\end{align}
the power spectrum for the thermal noise sound wave is
\begin{align}
	\langle \Pi_{\mathbf k} \Pi_{\bar {\mathbf k}} \rangle &= 
	(2\pi)^3 \delta(\mathbf k+\bar{\mathbf k}) P_\Pi(k),
	~
	P_{\Pi}(k)\equiv \frac{9}{8k^4  \beta \rho_0}.
\end{align}
The power spectrum of the velocity is obtained as
\begin{align}
	\langle \partial_t \Pi_{\mathbf k} \partial_t \Pi_{\bar {\mathbf k}} \rangle &= 
	(2\pi)^3 \delta(\mathbf k+\bar{\mathbf k}) P_{\partial_t \Pi}(k),
	~
	P_{\partial_t \Pi}(k)\equiv \frac{3}{8k^2 \beta \rho_0}.
\end{align}
Then the power spectrum of the density perturbation is
\begin{align}
	P_{\delta_\rho} = \frac{2}{\beta \rho_0}.
\end{align}
Thus, we find white noise.
Interestingly, the white noise does not depend on the viscosity $\eta$ and the physical time $t$.
Hence, energy stored in the Brownian sound waves is characterized only by the temperature and energy density of the environment.
In fact, 1-dimensional Brownian motion bounded in a harmonic potential was studied in Ref.~\cite{Chandrasekhar:1943ws}, and the author found that the squared displacement depends only on the mass, frequency, and temperature. The average of the squared displacement corresponds to $P_{\delta_\rho}$ in our theory.
$\beta \rho_0$ is understood as the entropy density of the universe.
Hence, the noise spectrum is determined by the inverse of the entropy density.
$P_{\delta_\rho}$ changes adiabatically as the universe expands.
Now, we restore the scale factor and set it to $a=1$ at present.
Hereafter $k$ is redefined back to comoving momentum.
The dimensionless spectrum is parameterized as
\begin{align}
	\frac{k^3}{2\pi^2}P_{\delta_\rho} = \frac{30 (k\beta)^3}{g_* \pi^4}.
\end{align}
We have $\bar \beta^{-1}=2.7$K for the cosmic microwave background today, which can be converted to $k\bar \beta \sim 2.7k\times 10^{-26}{\rm Mpc}$.
The observable anisotropy scale in the CMB is $k\lesssim 0.1{\rm Mpc}^{-1}$.
Hence, the density fluctuation that arises from the stochastic noise is $ 10^{-78}a^3$ at $0.1 {\rm Mpc}^{-1}$, which is negligibly small. 
The spectrum is $\mathcal O(1)$ for $k \beta \sim 1$.
The large noise can be a potential issue for perturbation theory.
In fact, the hydrodynamical approximation is valid for $(k/a) \beta = k\bar \beta \ll 1$, as the photon thermal distribution is given as $(e^{k\bar \beta}-1)^{-1}$.
This condition is recast into $k \beta \ll a$.
Hence, it does not make sense to extrapolate the spectrum to an extremely short scale since $a\ll 1$ in the early universe.
Thus, the hydrodynamical approximation fails before perturbation theory breaks down. 
%The behavior of $k\bar \beta \sim 1$ implies that the typical wavelength of thermal photons is $\bar \beta$, and thus, the fluid cannot be seen as a continuous object above the scale.

\section{Energy in Brownian sound waves}
\label{eib}

In the previous section, we considered the Brownian sound waves in a cosmological radiation fluid.
We found that the stochastic noise does not depend on the details of viscosity.
The spectrum is written as white noise, and the size of the noise at the observable scale in the CMB is tiny.
Although the stochastic noise does not directly affect the CMB anisotropy, there is an intriguing implication.

In the last section, we considered the cosmic expansion is slow, compared to the time scale of Brownian motion.
Meanwhile, we set $a=1$ again and work in the physical momentum and time.
Consider integral of motion~\eqref{mb:ue3} for $H\ll \gamma_k$:
\begin{align}
	\partial_t E_{\mathbf k,\mathbf k'}   = Q_{\mathbf k,\mathbf k'},\label{mb:ue4}
\end{align}
where we defined the Hamiltonian $E$ and heating rate $Q$ by 
\begin{align}
	E_{\mathbf k,\mathbf k'} &= \frac{\rho_0+P_0}{2}\partial_t k\Pi_{\mathbf k}\partial_t k'\Pi_{\mathbf k'} + \frac{(\rho_0+P_0)c_s^2}{2}k^2\Pi_{\mathbf k} k'^2\Pi_{\mathbf k'},
	\\
	Q_{\mathbf k,\mathbf k'} &=\frac{(\rho_0+P_0)(\gamma_k+\gamma_{k'})}{2} \partial_t k\Pi_{\mathbf k}\partial_t k'\Pi_{\mathbf k'} + \frac{(\rho_0+P_0)\partial_t k'\Pi_{\mathbf k'} k\xi_{\Sigma,\mathbf k}+ E\partial_t k\Pi_{\mathbf k} k'\xi_{\Sigma,\mathbf k'} }{2}.
\end{align}
$E$ corresponds to the energy of sound waves derived in Ref.~\cite{Landau1987Fluid} in a nonrelativistic setup, while our formulation applies to relativistic cases.
%In the absence of noise and dissipation, $Q=0$.
%In this case, $E$ is conserved, which corresponds to the generator of the diagonal time translation.
%Time translational symmetry is explicitly broken for $Q\neq 0$.
%
The long time limit of the stochastic average of $Q$ is found as zero: 
\begin{align}
	\lim_{t\gamma_k \to \infty}\langle  Q_{\mathbf k,\mathbf k'}\rangle=0,	
\end{align}
so that the net energy exchange between the sound wave system and the radiation environment is balanced.
This implies that the sound wave itself is in thermal equilibrium. %~\footnote{The explicitly broken time translation is restored in the long time limit. See Ref.~\cite{Hongo:2018ant} for a recent progress about time translational symmetry in open systems.}.
Energy stored in the $k$ mode is obtained as
\begin{align}
	\langle E_{\mathbf k,\mathbf k'} \rangle = (2\pi)^3 \delta(\mathbf k+\mathbf k')\frac{1}{2\beta}.\label{65}
\end{align}
When integrating Eq.~\eqref{65} over the momentum space, we encounter the UV divergence simply because we include infinitely short wavelength modes in the calculation.
As we discussed at the end of the last section, the hydrodynamical approximation is valid only for the $k\ll\beta^{-1}$.
Hence, we may put some cut-off scale of the integral as $\Lambda \beta^{-1}$ with $\Lambda \ll 1$.
In this case, we find
\begin{align}
	\rho_{\pi,\Lambda} = \frac{\Lambda^3}{12\pi^2 \beta^4}.
\end{align}
We cannot fix $\Lambda$ in the present theory.
Another method for the regularization would be introducing a thermal distribution for the sound wave.
The energy exchange between each $k$ mode and the environment may establish a thermal distribution for the whole sound wave system.
Eq.~\eqref{65} is classical, which may be understood as the small $\hbar$ limit of the quantum weight for bosons:
\begin{align}
	 \lim_{\hbar \to 0}\frac{\hbar k}{e^{\hbar k \beta}-1} = \frac{1}{\beta}.\label{rhodist}
\end{align}
This is a very ad-hoc assumption, but with this thermal distribution, we find a finite integral of \eqref{65}:
\begin{align}
	\rho_{\pi,{\rm th}} = \frac{\pi^2}{60\beta^4}.
\end{align}
Thermal distribution~\eqref{rhodist} has the peak around $k \beta\sim 1$ where the hydrodynamical approximation is not valid. 
This is a puzzle that thermalization process redistributes energy to the UV region where hydrodynamical approximation is not valid.
One intriguing observation is the energy density of all sound waves scales as $a^{-4}$ when restoring the scale factor~\footnote{When restoring the scale factor, one finds $\delta(\mathbf k+\mathbf k')\to a^{-3}\delta(\mathbf k+\mathbf k')$ and $\Lambda \to a\Lambda$. $\beta$ in Eq.~\eqref{rhodist} is rescaled to $\bar \beta$.}.
The energy contribution of Eq.~\eqref{rhodist} in the total energy density today is 
\begin{align}
	\Omega_{\pi,\rm th} = \frac{1}{4}\Omega_{\gamma 0},~\Omega_{\pi,\Lambda} = \mathcal O(\Lambda^3)\Omega_{\gamma 0}.
\end{align}
where $\Omega_{\gamma 0}\sim 10^{-5}$ is the energy fraction of radiation today.
The extrapolation to today may be too naive; however, in any case, the thermalized sound wave system effectively adds extra radiation to the background.
If we assume thermal distribution, the whole sound wave system corresponds to a half relativistic degree of freedom.

The energy contribution to the cosmic expansion is also read from the second-order component of $- T^0{}_0$.
Up to second order in the NG modes, we have
\begin{align}
	(K^{-1})_\alpha{}^\mu &= \delta_\alpha{}^\mu + \partial_\alpha \pi^\mu + \partial_\alpha \pi^\lambda \partial_\lambda \pi^\mu+\mathcal O(\partial \pi^3),
\\
	b^{-1} &= a^{-1} \left( 	1 - \partial_0 \pi^0 - \partial_0 \pi^i \partial_i \pi^0  
	+\frac{1}{2}\partial_0 \pi^i  \partial_0 \pi_i \right)+\mathcal O(\partial \pi^3).
\end{align}
Then
\begin{align}
	u^0 
	= a^{-1}\left( 	
	1    +\frac{1}{2}\partial_0 \pi^i  \partial_0 \pi_i
	\right),
	~
	u_0 = -a \left( 	
	1    +\frac{1}{2}\partial_0 \pi^i  \partial_0 \pi_i
	\right),
~	\gamma^0{}_0 =  - \partial_0 \pi^i  \partial_0 \pi_i.
\label{u0second}
\end{align}
The thermal radiation energy density in the rest frame~\eqref{def:thdens} is perturbed via the temperature:
\begin{align}
	\rho_0 = \frac{g_*\pi^2}{30a^4 \bar \beta^4}\left( 	1 - 4\partial_0 \pi^0 +6 \partial_0 \pi^0\partial_0 \pi^0 - 4\partial_0 \pi^i \partial_i \pi^0  
	+2\partial_0 \pi^i  \partial_0 \pi_i \right).
\end{align}
Carefully recovering the scale factor in the above calculation, the noise averages are obtained as
\begin{align}
	\langle \partial_0 \pi^i  \partial_0 \pi_i \rangle &= 
   a^{-4}\int \frac{d^3k}{(2\pi)^3} \frac{3}{8\bar \beta \rho_0},
   \\
   \langle \partial_0 \pi^0  \partial_0 \pi^0 \rangle &= 
   a^{-4}\int \frac{d^3k}{(2\pi)^3} \frac{1}{8\bar \beta \rho_0},
   \\
   \langle \partial_0 \pi^i \partial_i \pi^0 \rangle &=a^{-4} \int \frac{d^3k}{(2\pi)^3} \frac{3}{8\bar \beta \rho_0},
\end{align}
where we obtain $\pi^0$ by integrating $\partial_0 \pi^0$ as
\begin{align}
	\pi^0_{\mathbf k} = -\frac{k^2}{3} \int^{x^0} d\bar x^0 \Pi_{\mathbf k}(\bar x^0).
\end{align}
To summarize, one finds
\begin{align}
	\langle \rho_0\rangle = \frac{g_*\pi^2}{30a^4 \bar \beta^4} \left(1 + \mathcal O(\partial \pi^3)\right).
\end{align}
Similarly, $\langle P_0\rangle$ is unperturbed up to second order.
Hence, the back reaction comes from the perturbed 4-velocity~\eqref{u0second}, and we find
\begin{align}
	  \langle -T_{(2)}^0{}_0 \rangle = - (\rho_0+P_0)\langle (u^0u_0+1)\rangle = \frac{1}{2a^4}\int \frac{d^3k}{(2\pi)^3} \frac{1}{\bar \beta} .   
\end{align}
Thus, we reproduce Eq.~\eqref{65}.
Similarly, one can evaluate the back reaction to pressure and find 
\begin{align}
	  \langle \delta^j{}_iT_{(2)}^i{}_j \rangle = (\rho_0+P_0)\langle u^iu_i\rangle =  \langle -T_{(2)}^0{}_0 \rangle,   
\end{align}
which is the radiation equation of state.
The equation of state is independent of a prescription to the UV divergence.
Hence, the radiative scaling for the sound wave system is reproduced.

\section{Conclusions}
\label{conc}

Acoustic dissipation in the cosmic microwave background~(CMB) is explained in the Boltzmann theory.
Regarding the sound waves in the CMB as the linear response to the thermal radiation, the fluctuation-dissipation relation implies the existence of the noise counterpart.
Such stochastic noise is not incorporated into the established kinetic theory for cosmological fluids since it is a theory in the classical limit.
This missing piece may explain the inconsistencies in the Lambda CDM cosmology reported in the literature.
In this paper, we apply the effective field theory for fluctuating hydrodynamics to cosmological photon baryon plasma and find the stochastic noise inherent in the plasma.

We derived general formulas for fluctuation-dissipation relation in cosmological perturbation theory in a Friedmann Lemaitre Robertson Walker~(FLRW) background spacetime.
For simplicity, we considered the deep radiation dominant, where the photon-baryon plasma is dominated by photons.
In this setup, many EFT coefficients are simplified, and we found the noise on the sound waves.
Interestingly, the density power spectrum does not depend on the details of viscous coefficients in the long time limit.
The spectrum is determined by the entropy density of the universe.
In the case of the CMB, the noise is many orders of magnitude smaller than the primordial perturbations at the CMB anisotropy scale; therefore, the CMB anisotropy is not directly affected by the noise.

We also considered energy stored in the sound waves and its impact on cosmology.
We found that the energy density of the sound wave of individual $k$ mode is similar to the energy of a thermal harmonic oscillator in classical statistical mechanics.
Similar to the Brownian motion bounded in a harmonic potential, the energy exchange between the sound wave system and the environment is balanced in the long time limit.
We propose to understand that the sound waves themselves are in thermal equilibrium in this terminal state.
As different $k$ modes are excited in the present setup, we encounter the issue of UV divergence to evaluate the total energy density in the sound wave system.
The UV cut-off scale of hydrodynamical theory applies to the integral, and we found the energy density scales as $a^{-4}$, like radiation.
We also consider another regularization by introducing a thermal distribution for the NG modes.
The balanced energy exchange between the individual sound wave systems and radiation environment suggests the distribution of NG modes is thermal.
Then, we also found $a^{-4}$ scaling of the sound wave energy density.
In this case, the sound wave system carries a half relativistic degrees of freedom. 
We admit both regularization schemes are very ad hoc, so we cannot determine the actual size of the total energy in the sound wave system.
On the other hand, we found the radiation equation of state for the sound wave system by evaluating the back reaction to the energy density and pressure in the energy momentum tensor; hence, the conclusion about scaling is robust.
In any case, extra radiation species may have a big impact on the evolution of the universe and hence the fitting problem in the $\Lambda$CDM cosmology.
Further investigation of the energy in the sound waves will be necessary.

It is worth mentioning the similarity between our theory and various EFTs in cosmology. In the EFT of inflation~\cite{Cheung:2007st}, one can write the effective theory for cosmological perturbations as a nonlinear realization of diffeomorphism symmetry. 
Around the inflationary solution, one can write the action of NG modes associated with the time diffeomorphism and compute the initial curvature perturbations systematically based on the symmetry principles. 
In that case, the integration over the fast variables in Eq.~\eqref{changeval} is not assumed in the presence of the non-vanishing vacuum expectation value. 
The NG mode remains microscopic, so no hydrodynamical constraints are imposed. 
Also, the effective action is classical, i.e., the leading order in the advanced field expansion in the Schwinger-Keldysh action. The same arguments apply to the EFT of dark energy~\cite{Gubitosi:2012hu}. 
In Ref.~\cite{Hongo:2018ant}, the author and their collaborators extended the EFT of inflation for the Schwinger-Keldysh formalism. 
However, the NG modes are still microscopic fields. On the other hand, the EFT of large-scale structure~\cite{Carrasco:2012cv} is an effective theory for a hydrodynamical system. 
Their approach is based on the equation of motion, and action formulation from symmetry principles is not considered essentially because they are Eulerian. 
In this context, Ref.~\cite{Porto:2013qua} considered the Lagrangian formulation of the EFT of large-scale structure while they focus on the underlying kinetic theory. 
In our perspective, one can regard the Lagrangian displacement as the hydrodynamical variable and construct the hydrodynamical effective theory for this NG mode in the Schwinger-Keldysh formalism. 
This extension will be discussed in a separate paper.

Before closing this paper, we comment on several interesting extensions of this work.
The new technique will apply to other situations in cosmology, such as (perturbed) recombination, Big Bang nucleosynthesis, electron-positron pair annihilation, neutrino decoupling, or phase transition.
Rapid changes in the matter contents will introduce non-trivial EFT coefficients and, thus, noise.
Entropy production is defined from the unitarity condition~\cite{Glorioso:2016gsa}, and thus, we may evaluate thermal history more systematically.
This technique will be useful to compute deformation in the Planck distribution of the cosmic microwave background~\cite{Zeldovich:1969ff,Sunyaev:1970er} by generalizing the method in Ref.~\cite{Pajer:2012qep}.

\appendix

\section{Trace inequality}
\label{apti}

In the main text, we used the following trace inequality to show the second unitarity constraint.

\textit{Theorem---.}
Consider two complex $m \times n$ matrices $A$ and $B$ and their absolute values $|A|\equiv \sqrt{A^\dagger A}$ and $|A^\dagger|\equiv \sqrt{AA^\dagger}$.
Then
\begin{align}
    \left|{\rm Tr}[A^\dagger B]\right |^2\leq {\rm Tr}[|A||B|] {\rm Tr}[|A^\dagger||B^\dagger|].\label{th:1}
\end{align}
A proof is found in Ref.~\cite{baumgartner2011inequality}.
This theorem applies to finite-dimensional complex matrices.
We extend it to operators in Hilbert space without rigorous justification for simplicity. 
We use this theorem for $A=\hat \varrho$ and $B=U^\dagger(g_{(2)})U(g_{(1)})$. 
Then we find 
\begin{align}
	|{\rm Tr}[\hat \varrho U^\dagger(g_{(2)})U(g_{(1)})]| \leq 1, \label{myeq}
\end{align}
where we used $|\hat \varrho| =|\hat \varrho^\dagger| =1$ and $|U^\dagger(g_{(2)})U(g_{(1)})|=|(U^\dagger(g_{(2)})U(g_{(1)}))^\dagger|=1$.

\section{Dynamical KMS symmetry}
\label{apdy}

The local KMS transformation is defined for the path integral and explicit for the sources. 
In practice, similar constraints on the effective action $I_{\rm eff.}$ with respect to the dynamical variables are convenient.
In fact, we may read the hydrodynamical modes as external sources in the path integral with respect to the fast variable~\eqref{fastint}.
Setting $g_{(1)}=g_{(2)}=g$ in the effective action, one finds
\begin{enumerate}
	\item [1'] The conjugate condition:
	\begin{align}
	I^*_{\rm eff.}[\pi_{(1)},g; \pi_{(2)},g] = -I_{\rm eff.}[\pi_{(2)},g; \pi_{(1)},g]	
	\end{align}
	\item [2'] The unitarity conditions:
		\begin{align}
	I_{\rm eff.}[\pi,g; \pi,g] = 0,~ \rm Im~I_{\rm eff.} \geq 0.
	\end{align}
	\item [3'] The dynamical KMS symmetry:
\begin{align}
	\pi_{(1)} &\to \tilde \pi_{(1)}, 
	\\ 
	\pi_{(2)} &\to \tilde \pi^{(\beta)}_{(2)},
\end{align}
instead of the local KMS transformations \eqref{kmsg1} and \eqref{kmsg2}.
\end{enumerate}
Now, the KMS transformation is imposed for the dynamical variables.
By construction, these are necessary conditions for 1-3 but are not sufficient, which is discussed in more detail in Ref.~\cite{Crossley:2015evo}.

Let us find the dynamical KMS transformations for the NG modes in the Keldysh basis.
Setting $g_A$ to zero, we will see the KMS transformation for the dynamical variables such that Eq.~\eqref{kmsloc} is reproduced.
The time reversal transformations for the physical and fluid spacetime coordinates, $(\tilde x^0,\tilde x^i)=(-x^0,x^i)$ and $(\tilde \sigma^0,\tilde \sigma^a)=(-\sigma^0,\sigma^a)$, yield the time reversal of $\pi^\mu$
\begin{align}
	\tilde \sigma^{\mu} &= \tilde x^\mu - \tilde \pi^\mu(\tilde x),
\end{align}
which is recast into
\begin{align}
	\tilde \pi^0(\tilde x) &= -\pi^0(x),
	\\
	\tilde \pi^i(\tilde x) &= \pi^i(x).
\end{align}
The time reversal of $\pi_A^\mu$ is the same as $\pi^\mu$.
The nondiagonal imaginary time translation is generated by $\xi_A^\mu = i\hbar\beta^\mu$.
$\sigma^\mu(x)$ is a set of four scalar fields, so we find the following gauge transformation:
\begin{align}
	\delta_\beta \sigma^{\mu} &= i\hbar \pounds_{\beta}\sigma^{\mu}
	\notag 
	\\
	&= i\hbar \beta^\alpha \partial_\alpha \sigma^{\mu}
		\notag 
	\\
	&=  i\hbar \beta^\mu - i\hbar \beta^\alpha \partial_\alpha \pi^{\mu}.
\end{align}
Hence, the NG mode is transformed as
\begin{align}
	\delta_\beta \pi^\mu = i\hbar \beta^\mu - i\hbar \beta^\alpha \partial_\alpha \pi^{\mu}.
\end{align}
To summarize, the dynamical KMS transformation for $\pi^\mu$ is given as
\begin{align}
	\pi^0(x)  &\to -\pi^0(-x^0,x^i) + \mathcal O(\hbar),
	\\
	\pi^i(x)  &\to \pi^i(-x^0,x^i) + \mathcal O(\hbar\pi^2).
\end{align}
On the other hand, $\pi_A^\mu$ is nonlinear with respect to the nondiagonal imaginary time translation:
\begin{align}
	\delta_\beta \pi_A^\mu =  i\hbar \beta^\mu + \mathcal O(\hbar^2) =  i\hbar \bar \beta \partial_0 \pi^\mu+ \mathcal O(\hbar^2).
\end{align}
Then, the dynamical KMS transformation for the advanced NG mode is 
\begin{align}
	\pi_A^0(x^0,x^i) &\to - \pi_A^0(-x^0,x^i) - i\hbar \bar \beta \frac{\partial}{\partial(-x^0)} \pi^0(-x^0,x^i),
	\\
	\pi_A^i(x^0,x^i) &\to \pi_A^i(-x^0,x^i) + i\hbar \bar \beta \frac{\partial}{\partial(-x^0)} \pi^i(-x^0,x^i).
\end{align}

\acknowledgments

The author would like to thank Xiao-Han Ma, Jia-Rui Li, Dongdong Zhang for useful discussions.
This work was supported in part by the National Natural Science Foundation of China under Grant  No. 12347101.

\bibliography{sample.bib}{}
\bibliographystyle{unsrturl}

\end{document}